\documentclass[fleqn,usenatbib]{mnras}

\usepackage{newtxtext,newtxmath}
\usepackage{tablefootnote}
\usepackage{threeparttable}
\usepackage[T1]{fontenc}
\usepackage{adjustbox}

\DeclareRobustCommand{\VAN}[3]{#2}
\let\VANthebibliography\thebibliography
\def\thebibliography{\DeclareRobustCommand{\VAN}[3]{##3}\VANthebibliography}

\usepackage{graphicx}	% Including figure files
\usepackage{amsmath}	% Advanced maths commands

\newcommand{\trendone}[1][ ]       { $ -0.0 _ { -0.0001} ^ { +0.0000 }$~#1}
\newcommand{\trendtwo}[1][ ]       { $ -0.003 _ { -0.009 } ^ { +0.017 }$~#1}
\newcommand{\trendthree}[1][ ]     { $ 0.2 _ { -0.3} ^ { +0.5 }$~#1}

\newcommand{\omegatwo}[1][deg]     { $ 295.7 _ { -3.6} ^ { +6.6}$~#1}
\newcommand{\ecc}[1][ ]            { $ 0.300 _ { -0.009} ^ { +0.013}$~#1}
\newcommand{\tzero}[1][d]          { $ 1880.4 _ { -0.2 } ^ { +0.7 }$~#1}
\newcommand{\logK}[1][ ]           { $ 72.2_ { -1.0} ^ { +1.0 }$~#1}
\newcommand{\Ptwo}[1][d]           { $ 52.04 _ { -0.02} ^ { +0.02 }$~#1}

\newcommand{\massratio}[1][ ]      { $ 0.56 _ { -0.01} ^ { +0.03 }$~#1}
\newcommand{\imutual}[1][deg]      { $ 16.8 _ { -1.4} ^ { +4.2}$~#1}
\newcommand{\gtwo}[1][deg]         { $359.05 _ { -41.2} ^ { +12.0 }$~#1}
\newcommand{\aabsini}[1][R$_\odot$]  { $ 91.22 _ { -10.7 } ^ { +2.67 }$~#1}
\newcommand{\acsini}[1][R$_\odot$]  { $ 70.923 _ { -2.03} ^ { +0.98}$~#1}
\newcommand{\mcsini}[1][M$_\odot$]  { $  11.87 _ { - 2.9} ^ { + 0.9 }$~#1}
\newcommand{\masbini}[1][M$_\odot$]  { $ 9.21 _ { - 1.4} ^ { + 0.6}$~#1}

\title[TIC~470710327]{Planet Hunters TESS IV: A massive, compact hierarchical triple star system TIC~470710327}

\author[Eisner et al.]{N. L. Eisner$^{1,2}$\thanks{E-mail: nora.eisner@new.ox.ac.uk},
C. Johnston$^{3,2}$, % \orcid{0000-0002-3054-4135},
S. Toonen$^{4}$,
A. J. Frost$^{2}$, %0000-0001-9131-9970
S. Janssens$^{2}$, % 0000-0002-9758-4289
C. J. Lintott$^{1}$,
\newauthor
S. Aigrain$^{1}$,
H. Sana$^{2}$,
M. Abdul-Masih$^{5}$,
K. Z. Arellano-C\'{o}rdova$^{6,7}$,
P. G. Beck$^{7,8}$,
\newauthor
E. Bordier$^{5,2}$,
E. Cannon$^{2}$,
A. Escorza$^{5}$,
M. Fabry$^{2}$, %0000-0003-4200-7852
L. Hermansson$^{9}$, 
S.~B.~Howell$^{10}$,
\newauthor
G. Miller$^{1}$,
S. Sheyte$^{11}$ 
S. Alhassan$^{12}$,
E. M. L. Baeten$^{12}$,	
F. Barnet$^{12,13}$, %	Department of Mathematics, Frostburg State University, Frostburg, MD 21532
S. J. Bean$^{12}$,	
\newauthor
M. Bernau$^{12}$,	
D. M. Bundy$^{12}$,	
M. Z. Di Fraia$^{12}$,	
F. M. Emralino$^{12}$,
B. L. Goodwin$^{12}$,	
\newauthor
P. Hermes$^{12}$,	
T. Hoffman$^{12}$,	
M. Huten$^{12}$,	
R. Jan\'i\v{c}ek$^{12}$,	
S. Lee$^{12}$,	
M. T. Mazzucato$^{12}$,	
\newauthor
D. J. Rogers$^{12}$,	
M. P. Rout$^{12,14}$,	
J. Sejpka$^{12}$,
C. Tanner$^{12}$,	
I. A. Terentev$^{12}$
D. Urvoy$^{12}$
\\
$^{1}$Sub-department of Astrophysics, University of Oxford, Keble Rd, Oxford, United Kingdom\\
$^{2}$Institute of Astronomy, KU Leuven, Celestijnenlaan 200D, 3001 Leuven, Belgium\\
$^{3}$Department of Astrophysics, IMAPP, Radboud University Nijmegen, P. O. Box 9010, 6500 GL Nijmegen, the Netherlands\\
$^{4}$Anton Pannekoek Institute for Astronomy, University of Amsterdam, 1090 GE Amsterdam, The Netherlands\\
$^{5}$ European Southern Observatory, Alonso de C{\' o}rdova 3107, Vitacura, Casilla 19001, Santiago de Chile, Chile\\
$^{6}$ Department of Astronomy, The University of Texas at Austin, 2515 Speedway, Stop C1400, Austin, TX 78712, USA
\\
$^{7}$ Instituto de Astrof\'{\i}sica de Canarias, E-38200 La Laguna, Tenerife, Spain \\
$^{8}$ Institute of Physics, University of Graz, NAWI Graz, Universitätsplatz 5/II, 8010 Graz, Austria\\
$^{9}$ Sandvretens Observatory, Linn{\'e}gatan 5A, 75332, Uppsala, Sweden \\
$^{10}$ NASA Ames Research Center, Moffett Field, CA 94035, USA\\
$^{11}$ Institute of Astronomy and Astrophysics (IAA), Université libre de Bruxelles (ULB), CP 226, Boulevard du Triomphe, 1050 Bruxelles, Belgium \\
$^{12}$ Citizen Scientist, Zooniverse c/o University of Oxford, Keble Road, Oxford OX1 3RH, UK \\
$^{13}$ Department of Mathematics, Frostburg State University, Frostburg, MD 21532, USA \\
$^{14}$ The Rockefeller University, 1230 York Avenue, New York, NY 10065 USA
}

\date{Accepted for publication by MNRAS on 27 November 2021.}

\pubyear{2021}

\begin{document}
\label{firstpage}
\pagerange{\pageref{firstpage}--\pageref{lastpage}}
\maketitle

\begin{abstract}
We report the discovery and analysis of a massive, compact, hierarchical triple system (TIC 470710327) initially identified by citizen scientists in data obtained by NASA's Transiting Exoplanet Survey Satellite (\textit{TESS}). Spectroscopic follow-up observations obtained with the {\sc hermes}  spectrograph, combined with eclipse timing variations (ETVs), confirm that the system is comprised of three OB stars, with a compact 1.10 d eclipsing binary and a non-eclipsing tertiary on a 52.04 d orbit. Dynamical modelling of the system (from radial velocity and ETVs) reveal a rare configuration wherein the tertiary star (O9.5-B0.5V; 14-17~M$_{\odot}$) is more massive than the combined mass of the inner binary (10.9-13.2~M$_{\odot}$). Given the high mass of the tertiary, we predict that this system will undergo multiple phases of mass transfer in the future, and likely end up as a double neutron star gravitational wave progenitor or an exotic Thorne-{\.Z}ytkow object. Further observational characterisation of this system promises constraints on both formation scenarios of massive stars as well as their exotic evolutionary end-products. 
\end{abstract}



%%%%%%%%%%%%%%%%%%%%%%%%%%%%%%%%%%%%%%%%%%%%%%%%%%
%%%%%%%%%%%%%%%%% BODY OF PAPER %%%%%%%%%%%%%%%%%%

\section{Introduction}

% Compact hierarchical multiple systems, whereby the inner stellar binary is orbited by a third body with a significantly longer orbital period, often experience measurable dynamical interactions which can produce in exotic evolutionary scenarios. The

%- triple systems experience dynamcal interactiosn that manifest as ETVs, we see them in the RVs, %or the changes in the eccentricities/semi major axis in (KZ).
%
%- The evolution and the changes in the orbital characterisstics axan kead ti unique orbital %configureations. Rapid rotation, mergers, close compact systems and create common-envelope things %and exotic objects thoguht hte chnaging of the orbital properties/dynamical perurbatioson ot the %orbtial properties. 
%
%- OB stars exists excluisively in binary and multiple systems.
%
%- Triple stellar evolution is an important aspect of the evolution of massive stars from the %formation, evolution and death of .... 
%
%- In this paper we present... 
%

%massive stars are rare 
%we've found one 

%make the sentences flow more
%if we're considering a binary we need to consider this 
%strong pivot towards the triple
%whole new consideration if it's a compact hierarchical triple 
%fourth paragraph could go earlier. 
%put first WHY and then WHAT! 
%leave last paragraph where it is.
%removing linking words!!!
%We present a rare compact hierarchical triple star system comprised of three OB stars. 

%These systems are rare
Despite their intrinsic rarity implied by the initial mass function \citep[IMF; see, e.g., ][]{1955Salpeter,2010Bastian,2017Dib}, massive stars (M$\ge$8~M$_{\odot}$) provide radiative, dynamical and chemical feedback to their environment, driving evolution on a wide range of scales. The physical processes responsible for the formation, evolution, and death of massive stars, however, are not well understood \citep[e.g., ][]{2007Zinnecker,2014Tan}. The study of these processes is further complicated by the fact that such stars often have nearby stellar companions \citep{Sana2012}, which can affect their properties and evolution at all stages of their lives. Large scale spectroscopic, interferometric, and high contrast imaging surveys of OB stars have demonstrated that most, if not all, massive stars are formed in a binary or higher order multiple system \citep{Sana2013,Sana2014,Kiminki2012,Aldoretta2015,Moe17,MaizApellaniz2019,Rainot2020,Bodensteiner2021}.

The identification and characterisation of massive stars in multiple systems is crucial to discriminate between different formation and evolution scenarios, to place constraints on theoretical models and to understand the interactions and effects of multiple stars. \cite{Sana2012} and \cite{deMink2014} have already shown that the complex interactions between tides, angular momentum exchange, and stellar evolution in binary and higher order multiples fundamentally affect the evolution and thus the end product of nearly  70\%  of all early-type massive stars. In compact binary orbits, for example, processes including mass transfer, exchange of angular momentum and stellar mergers can open up new evolutionary pathways and end-products such as X-ray binaries, $\gamma$-ray bursts, stellar mergers, and gravitational wave events \citep{Sana2012,deMink2013}. The addition of a nearby third body further complicates the evolution. Interactions between three stars can induce different evolutionary pathways through von Zeipel-Kozai-Lidov cycles \citep[ZKL; ][]{1910Zeipel, Koz62,Lid62,Nao16,Ito2019}, or result in tertiary driven mass transfer that can lead to mergers, exotic common envelope systems, close double or triple degenerate systems, or contribute to the population of walk-away and run-away systems in our galaxy \citep{Antonini2017,Renzo2019,Stephan2019,Leigh2020,Hamers2021a,Glanz2021}.

Due to the high intrinsic brightness of massive stars, detecting and characterising non-eclipsing lower mass close companions using spectroscopy alone is challenging. Detailed characterisation of close companions, therefore, often relies on the detection of eclipses or other dynamical effects caused by the presence of a close companion, such as using radial velocity (RV) observations or observations of light travel time effects via eclipse timing variations (ETVs). Observations of the latter were made possibly with the advent of space-based photometric surveys with long time-base, high-precision, and high-cadence observations such as CoRoT, {\it Kepler}, {\it K}2 and \textit{TESS}. These space-based missions have enabled the detection of dozens of new triple and higher-order multiple systems through ETVs or multiply eclipsing events \citep{Conroy2014,Marsh2014,Borkovits2015,Rappaport2017,Hajdu2017,Sriram2018,2018Li,Borkovits2021}. Detailed modelling of such systems can lead to the determination of the absolute masses of the stellar components in some cases, or minimally the derivation of mass ratios, even in the absence of eclipses \citep{Borkovits2016}. 

Further considerations beyond light travel time effects need to be made for triple systems where the inner stellar binary is orbited by a third body with a longer period, known as a hierarchical triple system. Such a configuration can result in measurable dynamical perturbations to the orbit of the inner binary. If the orbital periods and the separations involved are short enough, these effects can be studied with a combination of ETVs (caused by light travel time effects and direct third-body perturbations) and/or RV observations. Measurements of both ETVs and RVs allows for the determination of precise stellar mass ratios and orbital parameters, including the mutual inclination between the inner and the outer orbit of the triple, which is thought to be indicative of the formation history.

In this paper we present a new compact, hierarchical triple system identified in {\it TESS} data consisting of one O- and two B-type stars. The system, which shows large ETV and RV variations, contains an inner $\sim1.1$~d eclipsing binary and a massive O9.5-B0.5V tertiary orbiting a common centre of mass in a $\sim52$~d orbit. The discovery of the system and the data are discussed in Sections~\ref{sec:target} - \ref{sec:spec_rv}. Section~\ref{sec:analysis} outlines the analysis of the photometric and spectroscopic data and Section~\ref{sec:discussion} discusses the system configuration, stability, formation and possible future evolution scenarios. Finally, the conclusions are presented in Section~\ref{sec:conclusion}.

\section{The target and its surroundings}
\label{sec:target} 
TIC 470710327 (BD+61 2536, TYC 4285-3758-1, V=9.6 mag, parallax=$1.06\pm0.24$ mas, distance=$950\pm220$ pc) was initially identified as an early B-type star by \citet{Brodskaya1953} and as a short period eclipsing binary with a period of 1.1047~d using photometric data obtained with the 0.25-m Takahashi Epsilon telescope in Mayhill, New Mexico, USA \citep{Laur2017}. The target is not a known member of a cluster or OB association \citep{Laur2017}. Two epochs of speckle interferometric measurements revealed a close companion at $\sim$ 0.5", with the position angle and distance of the companion advancing from $\theta=306.2$~deg and $\rho=0.533$" in 1987.7568 \citep{Hartkopf2000} to $\theta=303.7$~deg and $\rho=0.502$" in 2003.9596 \citep{Hartkopf2008}.

\begin{figure}
    \centering
    \includegraphics[width=0.45\textwidth]{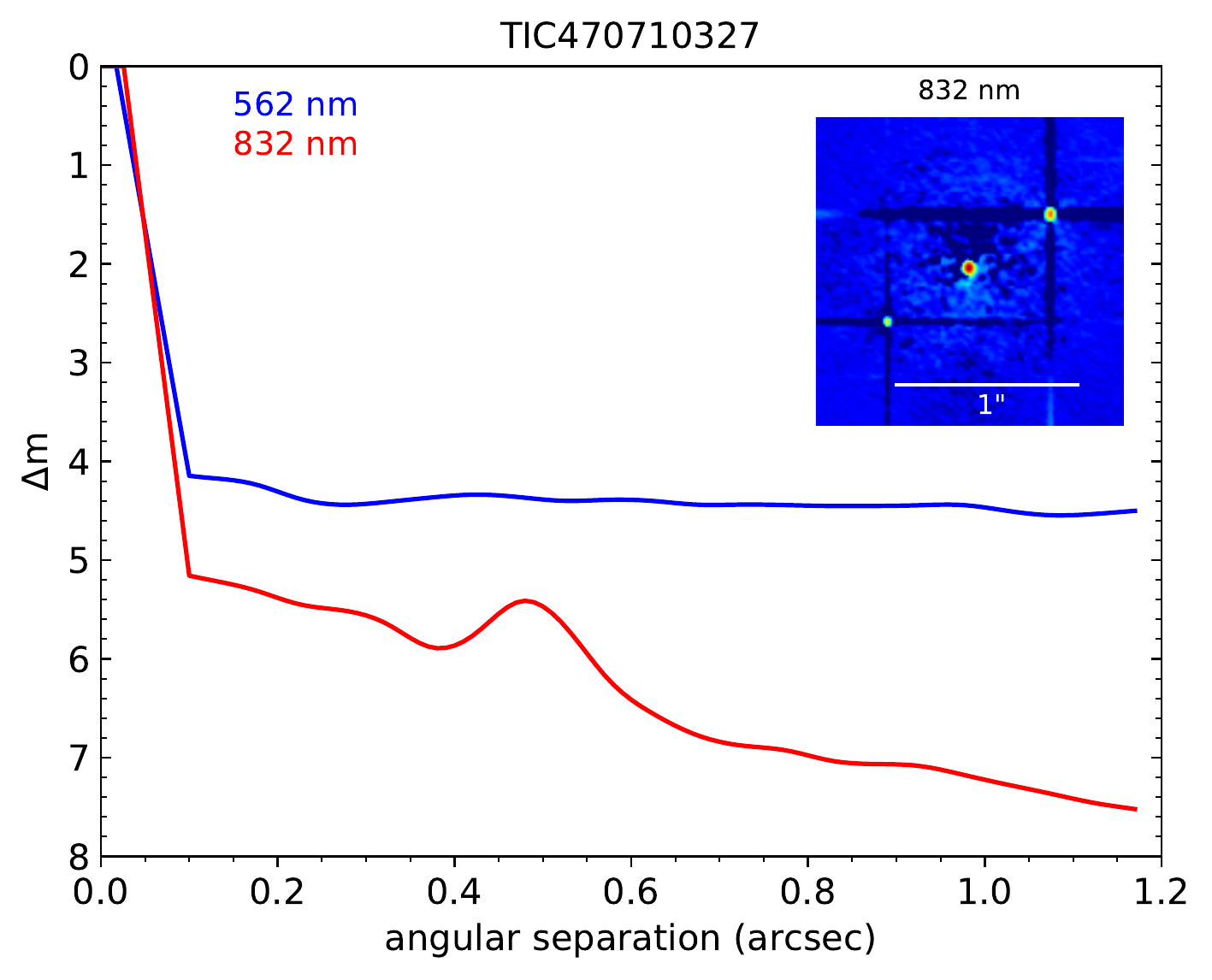}
    \caption{Contrast curves showing the 5 $\sigma$ detection sensitivity and the speckle reconstructed image for filters centred on 562 nm (blue) and 832 nm (red).}
    \label{fig:speckle}
\end{figure}

In order to calculate the magnitude differences between TIC 470710327 and the $\sim$ 0.5" companion, hereafter TIC 470710327$'$, we performed speckle imaging using the Zorro instrument on the 8.1-m Gemini South telescope on Cerro Pach\'{o}n, Chile \citep{Matson2019, Howell2011}. Observations were carried out on 15 August 2020 using the two-colour diffraction-limited optical imager with 60 msec exposures in sets of 1000 frames. The 5 $\sigma$ detection sensitivity and the speckle reconstructed image are shown in Figure~\ref{fig:speckle}. The data confirmed that the companion star is located at an angular separation of 0.529" with a position angle of 304.5~deg, which is in agreement with previous observations \citep{Hartkopf2000,Hartkopf2008}. The data showed that the companion has a magnitude difference of $\Delta$m=1.17~mag at 562 nm and of $\Delta$m=1.13~mag at 832 nm. 

To characterise the target's surroundings and to quantify the light contribution of nearby stars, we queried the early Gaia Data Release 3 catalog \citep[eDR3; ][]{gaiaedr3}. This search revealed a bright (V = 11.6, $\Delta$T$_{mag}$ =  2.4) nearby star at a separation of $\sim22$" (LS I +61 72), as well as a further six stars with a $\Delta$T$_{\rm mag}<$ 5 within 100" of the target (listed in Table~\ref{tab:nearby_obj}). The light contribution of these stars to our photometric observations will be discussed in  Section~\ref{subsec:phot_data}.

%The orange circles indicate the position of nearby stars that have a \textit{TESS} magnitude difference <5$\Delta {\rm T_{mag}}$ from the target star, as queried from the Gaia Data Release 2 catalog \citep{gaia2018gaia}. As the figure shows, there is only one other star that falls within the \textit{TESS} aperture (LS I +61 72) located at a separation of 21.84" and with $\Delta$T$_{\rm mag}$=2.4. Considering all sources with a $\Delta$T$_{\rm mag}<$ 5 within 100" of the target (listed in Table~\ref{tab:nearby_obj}), we expect TIC~470710327 (including the 0.5" companion) to contribute 88-89.5\% of the total light in the \textit{TESS} light curve. 

Overall, the triple system TIC 470710327 presented in this paper consists of a $P_{1}\sim$ 1.1 d eclipsing binary with a $P_{2}$ $\sim$ 52 d non transiting tertiary. Hereafter, the two stars in the inner, short period binary will be referred to as stars A and B, while the tertiary on the wide, outer orbit will be referred to as star C. The 0.5" companion star will be referred to as TIC 470710327$'$. Using the Gaia eDR3 list of sources in a 2$'$ radius to determine the local field density, and the magnitude contrasts given by the speckle observations, we show that the spurious association probability between TIC 470710327$'$ and TIC 470710327 is $\sim~1\times 10 ^{-5}$. This means that statistically there is an association between the target and the nearby companion. Assuming that the Gaia distance to this target \citep[$\sim950$ pc; ][]{Gaia2018} is accurate, an angular separation between TIC 470710327$'$ and TIC 470710327 of 0.5" corresponds to a physical separation of approximately 500 AU with an orbital period on the order of 1500-2000 years, depending on the mass of the companion. Following this, we will for the remainder of this paper assume that their physical separation is too large for TIC 470710327$'$ to have any dynamical effect on the triple system on the timescales of our observations. Furthermore, we note that both the astrometric excess noise (significant at  4000 $\sigma$) and the RUWE parameter (11.9) for TIC~470710327 are large. This is to be expected for unresolved multiple systems, but can contribute to an unreliable astrometric determination \citep{Ziegler2018,Lindegren2021}.  Conversely, the 22" companion has relatively small values for the RUWE parameter and astrometric excess noise, indicating that it is likely a single object.

\section{Photometric data and ETVs}
\label{subsec:phot_data}

\subsection{\textit{TESS}}

\begin{figure*}
    \centering
    \includegraphics[width=0.95\textwidth]{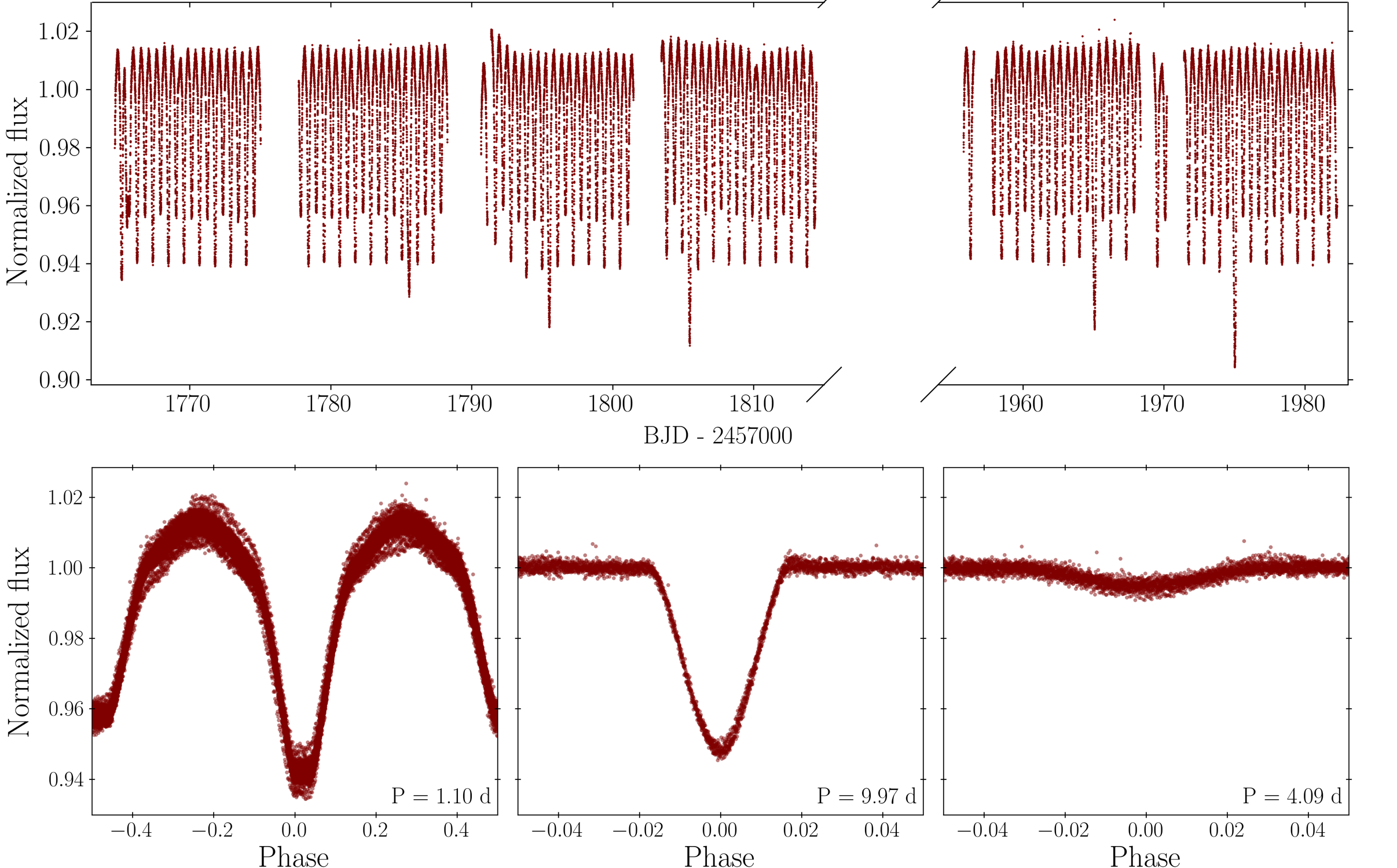}
    \caption{\textbf{Top}: full \textit{TESS} light curve of TIC 470710327 obtained during sectors 17,18 and 24. The dashed lines on the x-axis show a split in the axis. \textbf{Bottom}: light curve phase folded on $P_{1} =1.1$ d, $P_{3} =9.97$ d and $P_{4} =4.09$ d signals. Each panel represents one signal, where the other two signals had been masked out. Combined these three signals make up the light curve seen in the top panel. $P_{3}$ and $P_{4}$ are assumed to not be dynamically associated with the triple system presented in this paper, as discussed in Section~\ref{sec:analysis}.}
    \label{fig:fullLC}
\end{figure*}

TIC 470710327 was identified as a potential multiple system by citizen scientists taking part in the Planet Hunters TESS (PHT) citizen science project \citep{eisner2020method}. PHT, which is hosted by the Zooniverse platform \citep{lintott08, 2011Lintott}, has engaged nearly 30,000 registered volunteers in the search for planetary transit signals in 2-minute cadence light curves obtained by the Transiting Exoplanet Survey Satellite \citep[\textit{TESS}, ][]{ricker15}. In brief, each light curve is seen by 15 volunteers who identify times of transit-like events. Once a volunteer has classified a target, they are able to discuss the target on a discussion forum, and flag interesting systems to the PHT science team using searchable hashtags. TIC 470710327 was flagged as an interesting system on the discussion forum on 6th December 2019 \footnote{\url{https://www.zooniverse.org/projects/nora-dot-eisner/planet-hunters-tess/talk/2112/1195850?comment=2011869&page=1}}, due to the light curve containing multiple periodic signals: $P_{1} \sim 1.10$~d, $P_{3} \sim9.97$~d and $P_{4}\sim4.01$~d, as shown in the bottom panel of Fig.~\ref{fig:fullLC}. We note that for the remainder of this paper we assume that the stars associated with the $P_{3}$ and $P_{4}$ signals are not close enough to the triple to induce observable ETVs and that these signals do not originate from star C, as discussed further in Section~\ref{subsec:config}.

TIC~470710327 was monitored by \textit{TESS} in the 2-minute cadence data during Sectors 17, 18 and 24 of the nominal mission. The full \textit{TESS} data set, displayed in the top panel of Fig.~\ref{fig:fullLC}, spans $\sim$230 days, with the first subset covering nearly 60 days continuously, followed by a large gap of 140 days, before the second $\sim$30 day subset of observations. Visual inspection of the full \textit{TESS} light curve reveals a clear $\sim$1.10~d eclipsing binary system (Fig.~\ref{fig:fullLC}). The presence of full eclipses, combined with clear points of ingress and egress implies that the inner binary has a high inclination ($i_1$), close to 90 deg.

%Given the presence of the eclipses, we know that the inclination of this binary ($i_1$) needs to be above 70 deg and given the flat bottomed morphology we expect it to be closer to 90 deg.
%Due to the large pixel scale of \textit{TESS}, light contamination of nearby stars is common.

\begin{figure}
    \centering
    \includegraphics[width=0.45\textwidth]{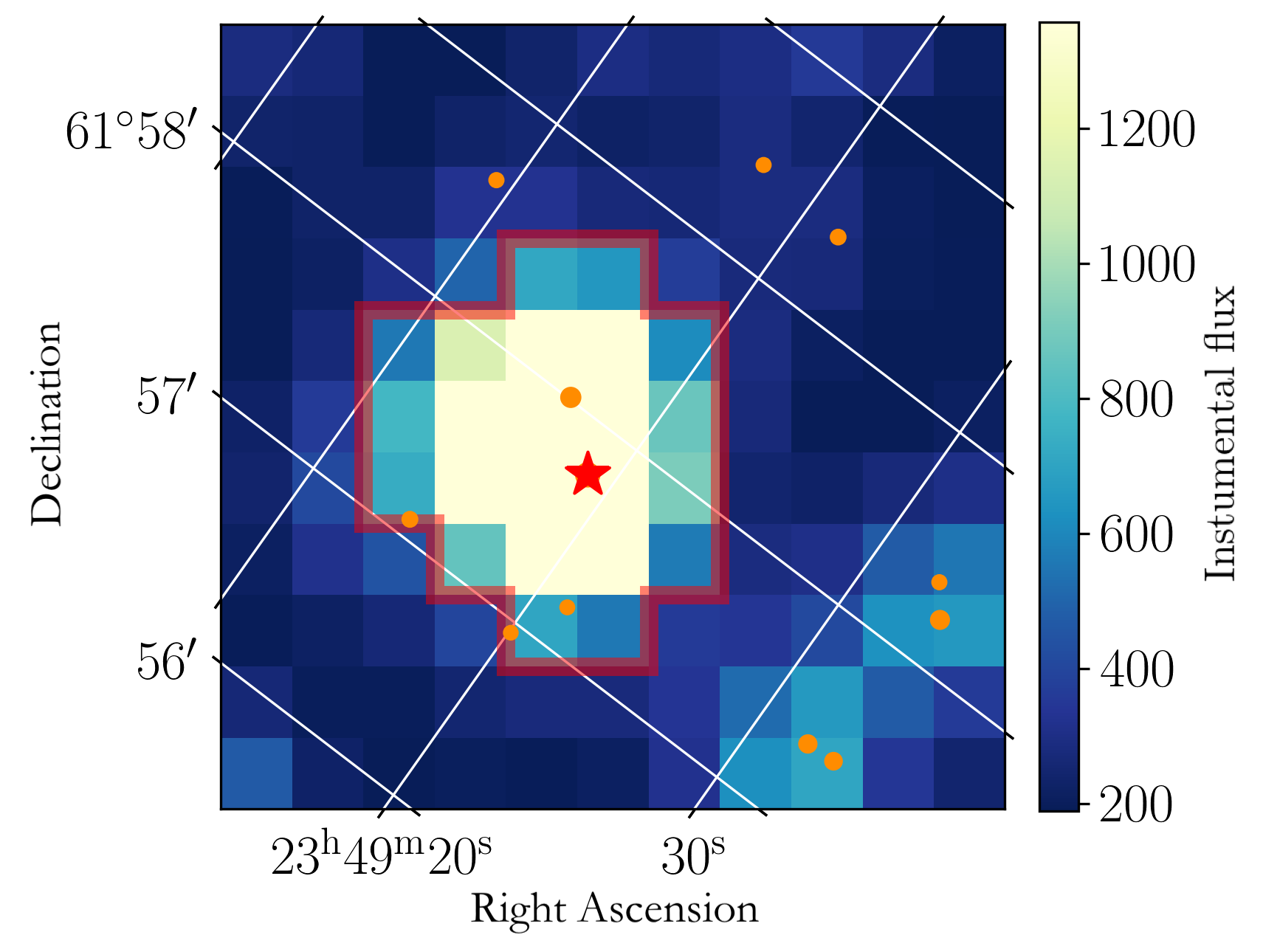}
    \caption{Average flux per pixel around TIC 470710327 obtained by \textit{TESS} during Sector 17. The orange dots show all neighbouring stars with Vmag<5$\Delta {\rm T_{mag}}$ from the target star, as queried from Gaia eDR3 \citep{gaiaedr3}. The red outline shows the \textit{TESS} aperture that was used to extract the flux in Sector 17.}
    \label{fig:nearbystars}
\end{figure}

The pixel aperture used to extract the light curve, as determined by the \textit{TESS} pipeline at the Science Processing Operations Center \citep[SPOC,][]{jenkins16}, is displayed as the red outline in Fig.~\ref{fig:nearbystars}. The orange circles indicate the position of nearby stars that have a \textit{TESS} magnitude difference $\Delta {\rm T_{mag}}$<5 from the target star, as queried from Gaia eDR3 \citep{gaiaedr3}. Considering all sources with a $\Delta$T$_{\rm mag}<$ 5 within 100" of the target (listed in Table~\ref{tab:nearby_obj}), we expect TIC~470710327 (including the 0.5" companion, TIC~470710327$'$ ) to contribute 88-89.5\% of the total light in the \textit{TESS} light curve. 

Similarly, using the magnitude contrast derived from the speckle imaging, we estimate that TIC~470710327$'$ contributes $\sim$26\% of the observed light at 832~nm. Thus, given the $\sim$10.5 - 12\% light contribution of the neighbour at 22", we estimate that TIC~470710327 and TIC~470710327$'$ contribute $\sim$65\% and $\sim$23\% of the total light observed by \textit{TESS}, respectively. As we show in Section~\ref{subsec:config}, the signals $P_{3} \sim9.97$~d and $P_{4}\sim4.01$~d cannot be hosted by the tertiary object in TIC~470710327, but likely originate from this composite contaminating light. For the remainder of the paper we will thus not consider their influence/contribution. Finally, we attempted to use both smaller apertures as well as centroid motions to determine the direction in which the photo-centre moved during the $P_3\sim9.97$ d and $P_4\sim4.01$ d eclipses. We found no discernible motion in the centroid positions and negligible change in eclipse depths when different apertures were used. This either suggests that all signals originate from a single \textit{TESS} pixel or that these signals have too small of a light contribution to result in a detectable centroid motion.

\subsection{Additional photometric observations}

In addition to the \textit{TESS} data, there are 531 archival photometric observations obtained with the 0.25-m Takahashi Epsilon telescope in Mayhill, New Mexico, USA between 2011 and 2013 \citep{Laur2017}; 108 photometric observations taken over 600 days by the SuperWASP-N camera located on La Palma, Canary Islands \citep{2006Pollacco}; as well as 1784 photometric observations obtained since 2012 by the ASAS-SN network of telescopes \citep{2014Shappee, 2017Kochanek}. We also obtained 456 photometric observations in the Johnson Cousins Bu band between 2020 March 21 and March 30, using a 0.41-m RC telescope with an Optical Guidance Systems and an SBIG STL-6303E camera, located at the Sandvreten Observatory, Sweden. Finally, we obtained 1415 photometric observations using the Las Cumbres Observatory (LCO) global network of fully robotic 0.4-m/SBIG and 1.0-m/Sinistro facilities. The LCO data were reduced and calibrated using the standard LCO Banzai pipeline. For both the LCO and the Sandvreten Observatory data we performed aperture photometry for TIC~470710327 and 5 comparison stars using the open source {\sc SEP} package \citep{2016Barbary, 1996Bertin}.

These archival and new photometric measurements significantly increased the baseline of the observations, allowing us to refine the period of the inner binary to $P_1$=1.104686$\pm$0.000004~d. The facilities used to obtain these measurements have significantly smaller pixel scales than \textit{TESS}, such that the extracted light curves do not include TIC~470710327$'$. This allowed us to confirm the stability of the dominant periodic signal, $P_1$=1.10~d, and verify that this binary signal does not originate from TIC~470710327$'$. Due to the sparse sampling of these five data sets we are unable to use these observations to investigate the stability or the origin of either the $P_{3}$ nor the $P_{4}$ signals. 

\begin{table}
    \centering
    \caption{Nearby GAIA eDR3 sources within 100" and $\Delta$Tmag$<$5 of the target.}
    \begin{tabular}{lccc}
        {\rm 2MASS Identifier} & Distance (arcsec) & $\Delta$~Tmag & $\Delta$~Vmag \\
        \hline
        23491896+6157459    & 0.000             & 0.00                & 0.00                \\
        23491667+6158004    & 21.845            & 2.40                & 2.35                \\
        23492132+6157124    & 37.395            & 4.76                & 5.75                \\
        23492010+6156568    & 49.643            & 4.82                & 5.07                \\
        23491426+6157050    & 52.642            & 4.38                & 5.64                \\
        23490933+6158367    & 84.892            & 4.65                & 5.36                \\
        23493217+6157231    & 95.863            & 3.35                & 3.69                \\
        23492162+6159230    & 98.898            & 4.45                & 5.11                \\
        % 23491758+6159263    & 100.989           & 4.69                & 5.12                \\
        % 23493273+6158224    & 103.697           & 4.67                & 6.18                \\
        % 23493339+6157236    & 104.112           & 3.64                & 3.67                \\
        % 23493361+6158140    & 106.991           & 2.99                & 3.23                \\
        \hline
    \end{tabular}
    \label{tab:nearby_obj}
\end{table}

\subsection{Eclipse Timing Variations}
\label{subsec:ETV}

Eclipse timing variations (ETVs), which are deviations from a strictly linear ephemeris of the eclipsing binary, can be used to detect or study additional gravitating bodies in a system. While the midpoint of eclipses of an isolated eclipsing binary are expected to occur at regular time intervals, dynamical perturbations from additional bodies, such as in a hierarchical triple system, can result in periodic deviations from the expected times of the eclipses. We searched for deviations from the predicted linear ephemeris of $P_1$, given by:
\begin{equation*}
\centering
    T_{\rm min, P / S} =  t_{0,P/S}+ 1.104686 \times (E),
\end{equation*}
where E is the cycle number since the reference orbit, $t_{0,P}$= 2458766.2700 is the reference epoch for primary eclipses, and $t_{0,S}$ = 2458766.82234 is the reference epoch for secondary eclipses. 

Using the \textit{TESS} data alone, we determined the deviations from this ephemeris following the methodology outlined by \citet{2018Li}. In brief, we determined the eclipse regions of the primary and secondary eclipses by extracting the minima of the second derivative around the time of the eclipses in the phase folded light curve. These translate to the phases of ingress and egress. Prior to this we masked the $P_3$=9.9733 d signal from the light curve. The lower amplitude signal from  $P_4$ was not removed due to the risk of introducing spurious signals. 

%We consider primary eclipses to occur at integer values of E and secondary eclipses to occur at half-integer values of E.
Next, we generated a model of the primary eclipse by fitting a trapezoid (the shape found to best represent the eclipses) to the smoothed, phase folded and subsequently binned light curve. This model was then fit to each individual primary transit where the only two free model parameters were the time of eclipse and the slope of an underlying linear trend. The latter was to allow for systematic effects that change the slope of the eclipse. The same methodology was independently carried out for the secondary eclipses. The individual fits to all eclipses, including both the primary and the secondary eclipses of the $P_1$=1.1047~d signal, were optimised using a Markov chain Monte Carlo (MCMC) approach, using the open source software {\sc exoplanet} \citep{exoplanet:exoplanet}. The observed minus calculated times of eclipse (O-C), which show variations on the range of $\sim \pm$ 6 minutes, are listed in Tables~\ref{tab:ETV_parimary} and \ref{tab:ETV_secondary} for the primary and secondary eclipses, respectively. The periodicity in the O-C curves, of P$_2\sim$52~d, is also observed in the radial velocities (Section~\ref{sec:spec_rv}), and is thus attributed to the tertiary star of the triple system, as discussed further in Section~\ref{sec:analysis}. 

%Upon inspection, the ETVs reveal a periodicity at P$_2$=52~d.

%\textbf{(From Ben and Peter)}

%Due to the complex nature of the system/area, there are no reported astrometric parallax measurements, to the best of our knowledge. Nonetheless, assuming that the star is a normal B2V star with an intrinsic brightness of Mv = -1.7 a dereddened observed magnitude of Vmag = 8, this would imply a photometric distance of ~870 pc for a single star and 1.2 kpc for a B2 binary with combined magnitude of 8 mag. 

%Adopting a mass of 7 MSun for the B2 star, a binary at an angular separation of 0.5 arcseconds would have an orbital period of around 3000 years.

% we only see the brightest star. Make it clear that eh brightest star is the tertiary and that it has to be the tertiary.

\section{Spectroscopic observations and radial velocity extraction}
\label{sec:spec_rv}

\begin{figure*}
    \centering
    \includegraphics[width=0.95\textwidth]{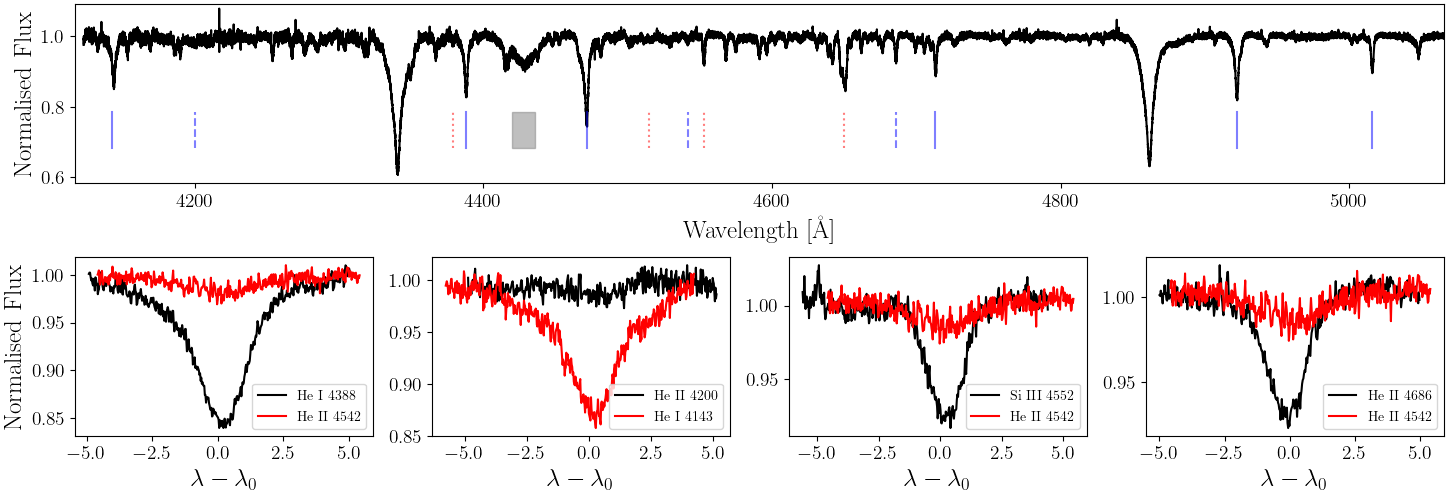}
    \caption{{\bf Top}: Shifted and median combined {\sc hermes} spectra of TIC 470710327~A (SNR > 80). Solid vertical blue lines denote He~I lines, dashed vertical blue lines denote He~II lines, dotted red lines denote metal lines, and the shaded grey region denotes a diffuse interstellar band. {\bf Bottom}: each panel displays a set of two lines used for diagnostic determination of spectral classification by \citet{sota2011}.}
    \label{fig:spectra}
\end{figure*}

We obtained 24 spectra between 31 January and 5 October 2020 using the {\sc hermes} spectrograph \citep[R$\sim$85\,000, ][]{Raskin2011} on the 1.2-m Mercator telescope at Observatorio del Roque de los Muchachos at Santa Cruz de la Palma, Canary Islands, Spain. The spectra were  reduced (extracted, order-merged, wavelength-calibrated) using the local {\sc hermes} pipeline \citep{Raskin2011}, and subsequently normalised using a spline fit \citep{AbdulMasih2021}. The spectra show strong He II 4686 lines and no sign of He II 4542 lines, indicating that the dominant signal originates from an early B star. The relative strengths of the He I, He II and Mg lines indicates that the dominant spectral contribution is consistent with that from an O9.5-B0.5V star \citep{sota2011}. The RV shifted and median combined spectrum is shown in Fig.~\ref{fig:spectra}, where the bottom panels show the relative depths of pairs of diagnostic spectral lines from \citet{sota2011} used to determine the spectral classification of the star. This spectral classification agrees with the previous estimates of the spectral class by \citet{Brodskaya1953} and \citet{Laur2017}. A spectral class of O9.5-B0.5V nominally corresponds to a 14-17~M$_{\odot}$ star \citep{Harmanec1988,Martins2005,Silaj2014}.

We searched for additional signals in the spectra using Least Squares Deconvolution \citep[LSD, ][]{Donati1997,Tkachenko2013}. In brief, this technique assumes that all lines in a spectrum have a common underlying  profile with varying depths depending on the particular line. This common profile is recovered by deconvolving the lines within a particular wavelength region with a line-list template with associated line-depths. This method was generalised by \cite{Tkachenko2013} to allow for multiple components in a spectra and to allow for each component to draw from a different line list. Here, we calculated LSD profiles of the {\sc Hermes} spectra over 4300-5200 \AA , using helium, carbon, nitrogen, and silicon lines whose rest wavelengths were computed from the Vienna Atomic Line Database (VALD-II, \citealt{Kupka1999VALD2}). We restricted ourselves to this range to minimize the potential noise contribution from weak lines. The resulting profiles did not reveal the presence of any additional components. The radial velocities of each spectrum were computed as the centre of gravity of the LSD profiles and are listed in Table~\ref{tab:RV}. Inspection of the RVs reveal the same periodic signal, of $P_2$=52~d, as seen in the ETVs (Section~\ref{subsec:ETV}).  As the $P_2$=52~d signal is seen in both the ETVs and the RVs, we confirm that the system is comprised of a $P_1$=1.1047~d eclipsing binary (stars A and B) and a $P_2$=52~d outer O9.5-B0.5V tertiary (star C) orbiting around a common centre of mass. 

\section{Joint ETV and RV modelling}
\label{sec:analysis}

In this section we describe the joint modelling of the ETV and the RV signals assuming that they are physically associated as a triple system. There are two main effects responsible for deviations from strict periodicity in eclipse timings: the light travel time effect (LTTE; geometrical contribution) and the dynamical effect. The former is a result of a change in projected distance from the centre of mass of the binary to that of the triple. The dynamical effect, on the other hand, results from physical changes in the orbit of the binary system due to the gravitational influence of the third body \citep{Borkovits2003}.

Modelling of the ETVs allowed us to derive properties including the mass ratio of the tertiary to the total mass of the system ($m_{\mathrm{C}}$/$m_\mathrm{{ABC}}$), the eccentricity of the tertiary ($e_{2}$), and the mutual inclination between the orbital plane of the binary and the orbital plane of the tertiary($i_{m}$). Following \citet{Borkovits2016},  perturbations produced by a close third body introduce deviations to a linear ephemeris according to:
\begin{equation}
\Delta = \sum_{i=0}^{3} c_{i} E^{i} + \left [\Delta_{\mathrm{LTTE}} +\Delta_{\mathrm{dyn}} \right ]_{0}^{E}.
\label{eqn:ETV}
\end{equation}
The first three terms multiplied by the cycle 
number $E$ represent corrections to the reference epoch ($c_0$), the 
orbital period ($c_1$), and any secular changes to the period ($c_2$).

The extent of the contribution of the LTTE to the perturbation depends on the light crossing
time of the relative orbit, $a_{AB}\sin i_2/\tilde{c}$ (where $\tilde{c}$ is the speed of light), as well as the configuration of the outer orbit:
\begin{equation}
\begin{aligned}
\Delta_{\mathrm{LTTE}} = -\frac{a_{AB} \sin i_{2}}{\tilde{c}} \frac{(1 - e_{2}^{2}) \sin (\nu_{2} + \omega_{2})}{1 + e_{2} \mathrm{cos}\nu_{2}},
\end{aligned}
\end{equation}
where $\nu_2$ is the true anomaly of the outer orbit, and is determined by $t_{0,2}$, P$_2$, and $e_2$. All quantities relating to the inner orbit ($P{_1}\sim$1.1 d) have a subscript 1, while all quantities relating to the outer orbit ($P_{2}\sim$52 d) have a subscript 2. Quantities 
with the subscript $AB$ refer to the individual components of orbit 1, whereas quantities with the subscript $C$ refer to the tertiary component in orbit 2. All symbols referring to a single quantity are explained in Table~\ref{tab:modeling}. 

The dynamical perturbation has a more complex dependence on the mass ratio
of the system, the ratio of the periods, as well as the mutual inclination 
of the two orbits, denoted as $i_m$. This term is given by:
\begin{equation}
\begin{aligned}
\label{eq:dynamical}
\Delta_{\mathrm{dyn}} = \frac{3}{4 \pi} \frac{m_{C}}{m_{ABC}} \frac{P_{1}^{2}}{P_{2}} (1-e_{2}^{2}) ^{-3/2} \\
\times \left [ \left (  \frac{2}{3} - \mathrm{sin}^{2}i_{m}  \right ) \mathcal{M}  + \frac{1}{2} \mathrm{sin}^{2}i_{m} \mathcal{S} \right ] \\
\end{aligned}
\end{equation}
with 
\begin{equation}
\begin{aligned}
\mathcal{M} = 3e{_2}\mathrm{sin} \nu_{2} - \frac{3}{4} e{_2}^{2} \mathrm{sin} 2\nu_{2} + \frac{1}{3}e{_2}^{3} \mathrm{sin} 3\nu_{2}
\end{aligned}
\end{equation}
and
\begin{equation}
\begin{aligned}
\mathcal{S} = \mathrm{sin}(2\nu_{2} + 2g_{2}) + e_{2} \left [ \mathrm{sin}(\nu_{2} + 2g_{2}) + \frac{1}{3} \mathrm{sin}(3\nu_{2} + 2g_{2})  \right ].
\end{aligned}
\end{equation}
All symbols are explained in Table~\ref{tab:modeling}. 

The RV variations of the tertiary component are given by: 
\begin{equation}
\begin{aligned}
V = \gamma + \frac{2 \pi a_{C} \mathrm{sin} i_{2}}{P_{2} \sqrt{(1 - e_{2}^{2})}} [e_{2} \mathrm{cos}(\omega_{2}) + \mathrm{cos}(\nu_{2} + \omega_{2})].
\label{eqn:RV}
\end{aligned}
\end{equation}
Here, all terms have the same subscripts as in the ETV equations, and the 
systemic velocity is given by $\gamma$. Given the overlap in parameters
between the ETV and RV models, as well as the complementary information held 
in the independent data sets, we are able to constrain the systems to a high degree. 

Model optimisation was carried out using a No U-Turn Sampling \citep[NUTS; ][]{2011Hoffman} Hamiltonian Monte Carlo (HMC) approach. In short, HMC is a class of Markov Chain Monte Carlo (MCMC) methods used to numerically approximate a posterior probability distribution. Whereas traditional MCMC techniques use a stochastic walk to explore a given $n$-dimensional parameter space, the NUTS algorithm makes use of a Hamiltonian description of probability distribution in order to more directly sample the posterior probability distribution of a set of model parameters $\theta$ given by Bayes' theorem: $p\left(\theta|d\right)\propto p\left(d|\theta,\sigma\right) \times p\left( \theta\right)$. The likelihood term $p\left(d|\theta,\sigma\right)$ is the evaluation of how well the model represents the data $d$ given the parameters $\theta$ and uncertainties $\sigma$. In our application, the model is given by the ETVs and RV variations in Eqns.~\ref{eqn:ETV} and \ref{eqn:RV}, such that:
\begin{equation}
    p\left(d|\theta,\sigma\right) \propto \left(\mathcal{M}_{ETV} - Y_{ETV} \right) + \left(\mathcal{M}_{RV} - Y_{RV} \right),
\end{equation}
where $\mathcal{M}$ refers to the model and $Y$ refers to the data. 

\begin{figure*}
    \centering
    \includegraphics[width=0.95\textwidth]{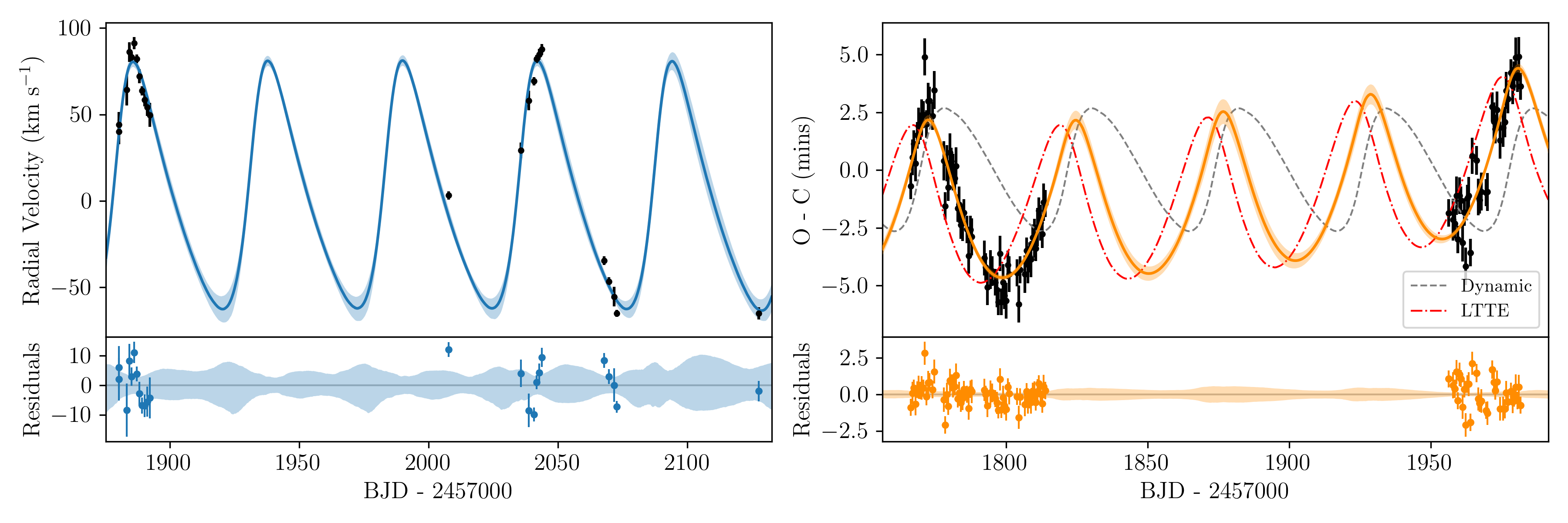}
    \caption{Joint MCMC model of the {\sc HERMES} RV data (left panel) and the extracted eclipse timing variations (right panel). The model parameters are presented in Table~\ref{tab:params}. The bottom panels show the residuals of the best fit. The overall ETV fit is a linear addition of the dynamical effect (small dashed grey line) and the light travel time effect (large dashed red line).}
    \label{fig:etvs_rvs}
\end{figure*}

In order to best exploit the complementary and overlapping information in the RV and ETV data, the two data sets were modelled jointly. This allowed us to simultaneously fit for, and better constrain, parameters that appear in both models (see Table~\ref{tab:modeling}). The joint analysis made use of the open source software packages {\sc exoplanet} and {\sc pymc3} \citep{exoplanet:exoplanet, exoplanet:pymc3}. The optimal parameter values and their uncertainties were calculated as the median and 67.8\% highest posterior density of the marginalised posterior distributions. The priors and extracted values for all sampled parameters are given in Table~\ref{tab:modeling}. The best-fitting model, as constructed from the values in Table~\ref{tab:modeling}, for the RVs and ETVs are shown in the left and right panels of Fig.~\ref{fig:etvs_rvs}. We note that the residuals in the ETVs and RVs show no evidence for additional periodicities.

% ADD A PARAGRAPH SAYING WHAT WE HAVE FOUND AND THAT IT IS A HIERARCHICAL TRIPLE SYSTEM WITH MASSIVE STARS. 

%The 9.9 d and 4.1 d signals show no obvious ETVs. 
\renewcommand{\arraystretch}{1.4} % Default value: 1
\begin{table*}
\centering
  \caption{System parameters either sampled or derived from the HMC optimisation, as well as parameters
            estimated from the {\it TESS} light curve. \label{tab:modeling}}  
  \begin{tabular}{lccccr}
  \hline
\textbf{Parameter}	&	\textbf{Symbol}	&	\textbf{Prior}	&	\textbf{Value} 	&	\textbf{Units} &	\textbf{Model} \\
\hline
\textbf{Sampled parameters} 	&		&		&		&	\\
Orbital period tertiary	&	$P_{2}$	&	$\mathcal{N}[52.1,2]$	&	\Ptwo[]	&	days  & ETV + RV \\
Semi-major axis, binary to COM	&	$a_{\mathrm{AB}} \sin i_{2}$	&	$\mathcal{U}[10,500]$	&	\aabsini[]	& R$_\odot$	& ETV \\
Tertiary to total mass ratio  ($m_{\mathrm{C}}$ / $m_{\mathrm{ABC}}$)	&	$q_{\mathrm{tot}}$	&	$\mathcal{N}[0.58,0.14]$	&	\massratio[]	&	& ETV \\
Tertiary eccentricity	&	$e_{2}$	&	$\mathcal{U}[0,0.4]$	&	\ecc[]	&	& ETV + RV \\
Binary eccentricity	&	$e_{1}$	&	--	&	0	&	& fixed \\
Observed argument of periastron	&	$\omega_{2}$	&	$\mathcal{U}[0,360]$	&	\omegatwo[]	& deg & ETV + RV \\
Mutual inclination	&	i$_{m}$	&	$\mathcal{U}[0,360]$	&	\imutual[]	& deg	& ETV \\
Dynamical argument of periastron	&	$g_{2}$	&	$\mathcal{U}[0,360]$	& \gtwo[]	&	deg & ETV \\
Semi-amplitude	&	$K$	&	$\mathcal{N}[81,20]$	&	\logK	&	& RV \\
$t_{0}$ of tertiary 	&	$t_{0,2}$	&	$\mathcal{N}[1878,25]$	&	\tzero[]	& BJD - 2457000	& ETV + RV  \\
Correction to $T_0$	&	$c_{0}$	&	$\mathcal{N}[0,5]$	&	\trendone[]	&	& ETV  \\
Correction to $P_1$	&	$c_{1}$	&	$\mathcal{N}[0,0.5]$	&	\trendtwo[]	&	& ETV  \\
Secular change to $P_1$ &	$c_{2}$	&	$\mathcal{N}[0,0.0001]$	&	\trendthree[]	&	& ETV  \\
\hline
\textbf{Derived parameters}	&		&		&		&	\\
Semi-major axis, tertiary to COM	& $a_{\mathrm{C}} \sin i_{2}$ &  --	&	\acsini[]			& R$_\odot$	& RV \\
Project mass of binary	    &	 $m_{\mathrm{AB}}\sin^{3}i_{2}$ &  --	& 	\masbini[]				& M$_\odot$	& ETV + RV  \\
Project mass of tertiary	&	 $m_{\mathrm{C}}\sin^{3}i_{2}$	 & -- & \mcsini[]			& M$_\odot$	& ETV + RV  \\
% True anomaly	&	$\nu_{2}$	&	--	&	--	&	& ETV + RV  \\
\hline
\textbf{Light curve extracted parameters}	&		&		&		&	\\
Orbital period binary	&	$P_{1}$	&	--	&	1.104686$\pm$0.000004	& days	& -- \\
$t_{0}$ of primary eclipse & $t_{0,1}$	&	--	& 1785.533	& 	BJD - 2457000 	& -- \\
\hline
\textbf{SED parameters}	&		&	   	&		&	\\
Mass of binary star A   &	$M_{A}$	&   --  &     6 - 7      & 	M$_{\odot}$  	& SED \\
Mass of binary star B   &	$M_{B}$	&   --  &     5.5 - 6.3      & 	M$_{\odot}$  	& SED \\
Mass of tertiary star C      &	$M_{C}$	&   --  &     14.5 - 16      & 	M$_{\odot}$  	& SED \\
Luminosity of binary star A   &	$L_{A}$	&   --  &      3.00 - 3.27      & 	log L/L$_{\odot}$  	& SED \\
Luminosity of binary star B   &	$L_{B}$	&   --  &     3.02 - 3.14      & 	log L/L$_{\odot}$  	& SED \\
Luminosity of tertiary star C      &	$L_{C}$	&   --  &     4.71 - 4.86      & 	log L/L$_{\odot}$  	& SED \\
Distance&	$d$	    &   --  &   4.0 - 4.5  	& 	kpc 	& SED \\
Reddening &	E(B-V)	&	--  &	0.40 - 0.44	& 	& SED \\
    \hline
   \noalign{\smallskip}
  \end{tabular}
%   \begin{tablenotes}\footnotesize
%   \item \textit{Note} -- $\mathcal{U}[a,b]$ refers to uniform priors between $a$ and $b$, $\mathcal{N}[a,b]$ to Gaussian priors with median $a$ and standard deviation $b$, and $\mathcal{F}[a]$ to a fixed value $a$.  
%   Inferred parameters and errors are defined as the median and 68.3\% credible interval of the posterior distribution. Subscript 1 revers to the inner binary system and subscript 2 to the outer tertiary orbiting around the inner binary. 
% \end{tablenotes}
\label{tab:params}
\end{table*}

% \subsection{SED fitting}
% \label{subsec:sed}
% {\bf @SOETKIN}

\section{System configuration, stability, and evolution}
\label{sec:discussion}

\subsection{Configuration}
\label{subsec:config}

\begin{figure}
    \centering
    \includegraphics[width=0.45\textwidth]{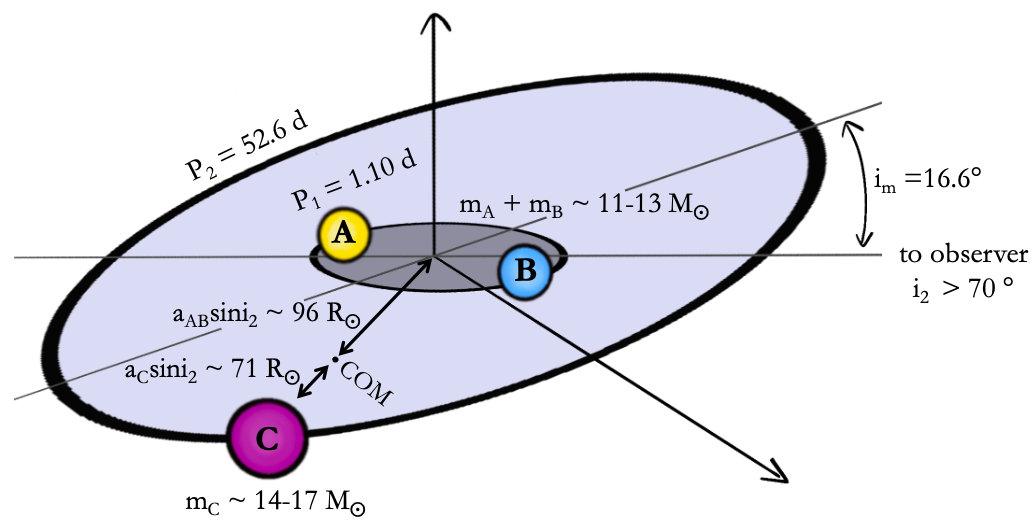}
    \caption{Schematic of the triple system. Relative sizes of the orbits are not to scale and for simplicity the orbits are depicted as circular.}
    \label{fig:schematic}
\end{figure}

The combination of radial velocity and photometric data revealed a dynamically interacting triple system, comprised of a close 1.1047~d eclipsing binary (stars A and B) with a massive companion on a wide, non-eclipsing, $\sim52$~d orbit (star C). From the dynamical modelling (presented in Section~\ref{sec:analysis}) we derived a mass ratio of the inner binary to the tertiary of $q=m_{AB}/m_C = a_C\sin i_2 / a_{AB}\sin i_2=0.7776$. Given the spectral classification of star C, of O9.5V-B0.5V, the tertiary star has a mass in the range of 14-17~M$_{\odot}$, meaning that the combined mass of stars A and B is in the range of 10.9-13.2~M$_{\odot}$. By considering both the spectroscopic mass range of the tertiary and the estimated mutual inclination $i_m$, we can derive limits on the inclination of both the orbit of inner binary $i_1$ and the orbit of the tertiary $i_2$ on the sky. From these considerations, we find that $i_1$ lies in the range: $77.9^{ \circ } \le i_1 \le 90^{ \circ }$ and $i_2$ lies in the range: $62.5^{ \circ } \le i_2 \le 71.1^{ \circ }$. The high inclination of $i_1$ agrees with the observed flat bottom, i.e. full, eclipses observed for the short 1.1~d orbit. Conversely, from the presence of flat-bottomed eclipses in the TESS lightcurve, we can infer that the inner binary has a large inclination: $i_1 >= 80^{\circ}$. By combining this inferred range with the mutual inclination derived from the ETV modelling, we arrive at an estimate for the inclination of the tertiary on the sky: $i_2 \in 59^{\circ} - 74^{\circ}$. Using this, we can infer physical ranges for $m_{AB}\in8.8~M_{\odot}-15.6 $ and $m_c\in 10.1~M_{\odot} - 20.3~M_{\odot}$. This wide range is in agreement with the mass ranges inferred from the spectral type of the tertiary star.
An overview of the triple system is presented in Fig.~\ref{fig:schematic}. We note that while the triple and the 0.5" companion star, TIC 470710327$'$, are statistically associated, the companion is not assumed to have a detectable effect on the observed dynamics of the triple given its separation and short period covered by the observations.

In addition to the eclipsing $P_1$=1.1047~d binary signal, the {\it TESS} light curve contains two further periodic eclipsing signals with $P_3$=9.9733~d and $P_4$=4.092~d. The bottom panel of Fig.~\ref{fig:fullLC} displays the {\it TESS} light curve phase folded on $P_3$ (middle panel) and $P_4$ (right panel). We note that due to the lack of visible `secondary' eclipses, we cannot distinguish between periods of $P_3$ and $P_4$ or twice those values. We can, however, determine that all three sets of eclipses are of different objects, as the morphology of all of the eclipses are constant in time, and points of overlap between the 1.10~d and 9.97~d signal are reproduced as linear additions of the different eclipse signals. Furthermore, we see no dynamical evidence in the RVs or the ETVs of either the 9.97 d or the 4.09 d signals being part of the same system. Using Eqn.~\ref{eq:dynamical}, we can show that a dynamically interacting body with $P_3$=9.97~d would produce an amplitude in the ETVs that is at least $(9.97/1.10)^2 = 82$ times larger than the dynamical amplitude induced by the $P_2$=52~d tertiary. We see no evidence of this signal in the data. Using the same argument we also find no evidence of the $P_4$=4.09~d signal being associated with the tertiary star of the triple. We note that $P_1$=1.10~d and $P_3$=9.97~d is close to a 1:9 resonance ($P_3$/$P_1$ = 9.028). While this could indicate that the signals are related, we do not have sufficient data at present to further investigate their association. 

As noted above, the presence of two additional bright sources in the {\it TESS} aperture, at separations of 0.5" and 22", overall contributing over $\sim$10\% of the light in the aperture, make it difficult to determine the exact source of all of the periodic signals seen in the \textit{TESS} light curve. Through the RVs and ETVs, however, we can associate the triple system with the brightest O9.5V-B0.5V star in the data. This is further corroborated by the data obtained with the Takahashi Epsilon telescope, which has a pixel scale of 1.64", which allowed us to confirm that the $P_1$=1.10~d signal lies on the O9.5V-B0.5V target star and not the 22" companion star that lies within the same \textit{TESS} aperture.

% Due to the sparse sampling of this archival data set, we are unable to recover the P$_3$=9.97 d or P$_{4}$ $\sim4.41$~d signals in the Takahashi Epsilon telescope observations. 

%\begin{figure}
%    \centering
%    \includegraphics[width=0.45\textwidth]{Figures/phase_folde%d_long_period.png}
%    \caption{TESS light curve phase folded at periods of %$P_{4} = 9.9733$ d (top panel) and  $P_{3} =4.1$ d (bottom %panel). The signal of the inner binary,  $P_{1} =1.1$ d %was removed prior phase folding.}
%    \label{fig:additional_phot_signals}
%\end{figure}

\subsection{SED Modelling}
\label{subsec:sed}
We investigate the properties of the components of this system via grid-fitting analysis of the composite spectral energy distribution (SED) of TIC~470710327 and TIC~470710327$'$, using photometric data from Vizier (Table~\ref{table_data_vizier}). In order to reduce the degeneracies present in composit SED modelling, we use the synthetic SEDs from a grid of MIST isochrone models \citep{choi2016}. This enforces that all components of TIC~470710327 and TIC~470710327$'$ are the same age and located at the same distance, and ensures realistic physical parameters  for all components. The models are reddened according to $F_{\text{red}} = 10^{-A_{\lambda}/2.5}F_{\text{mod}}$ and $A_{\lambda} = E(B-V)R_{\lambda}$, where E(B-V) is a free parameter, and $R_{\lambda}$ is calculated according to \citet{Cardelli1989}. Each model is evaluated against the data using a $\chi^2$ metric.

We fit a composite four-component SED to the observations, assuming one component for each member of the triple system, and one for the nearby 0.5" companion, TIC 470710327$'$. The light contributions derived in Section~\ref{subsec:phot_data} are used to inform the model, such that the total light contribution of the triple system is fixed to 77\% in the V-band, with the remaining 23\% originating from TIC 470710327$'$. Additionally, we enforce that the luminosity of the primary component of the close binary must be higher than that of the secondary, i.e., $L_A > L_B$. Furthermore, we adopt the spectroscopically derived mass range for the tertiary: $M_C\in 14~M_{\odot}-17~M_{\odot}$ and only accept solutions that satisfy the triple system mass-ratio of $q_c=0.776$. This results in 1019 possible solutions, which we rank according to their $\chi^2$.

\begin{figure}
    \centering
    \includegraphics[width=0.99\columnwidth]{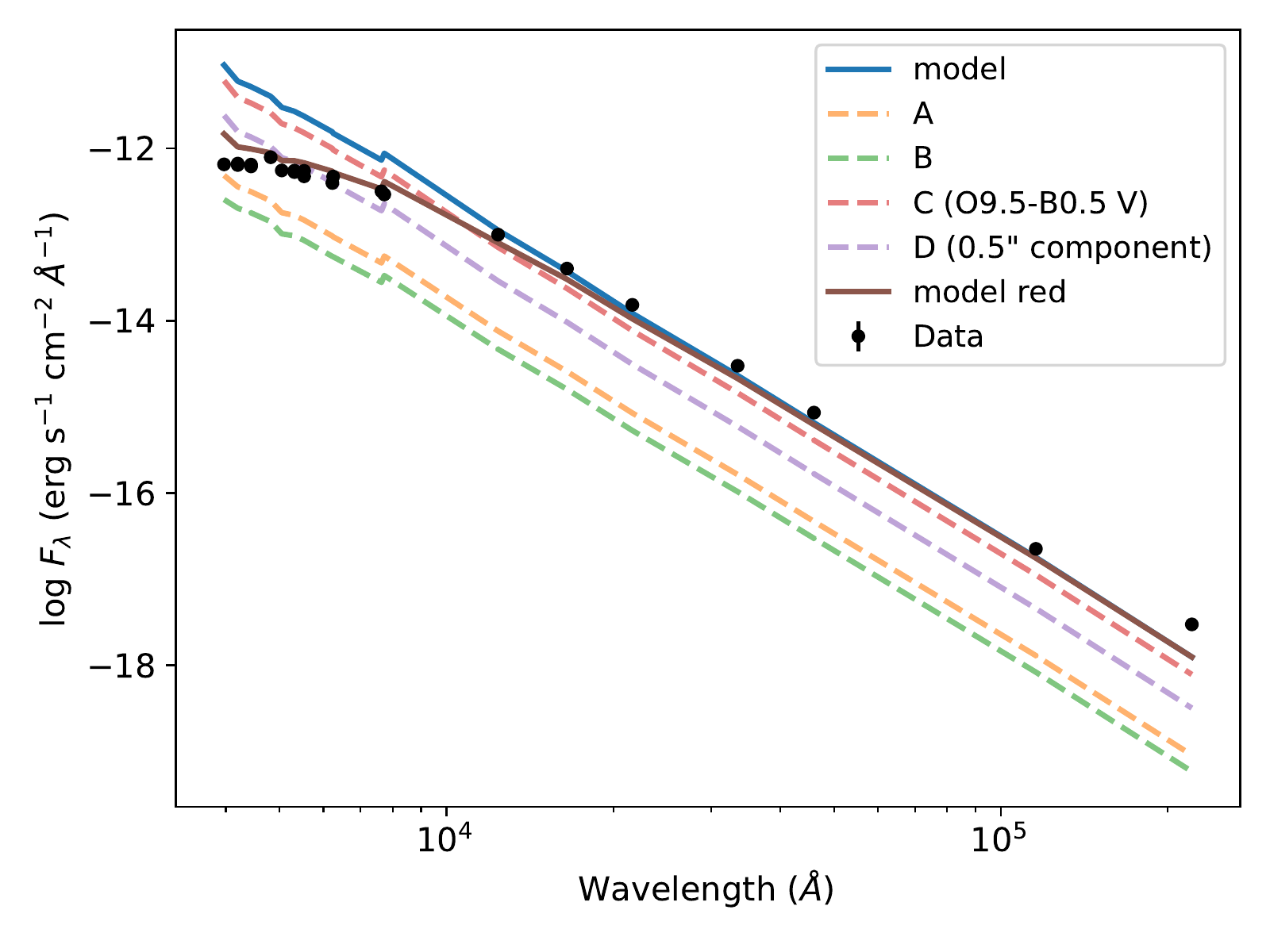}
    \caption{Observed SED (black points) with composite and individual model contributions for the best fitting composite model. The individual contributions are un-reddened, whereas we plot both the reddened (maroon) and un-reddened (blue) composite models.}
    \label{fig:sed}
\end{figure}

The solutions favour masses of the OB star of $M_C\in 14.5-16~M_{\odot}$, whereas the solutions for the components of the close binary orbit favour a mass ratio $q_1\in0.9-1.0$ with masses for the primary components ranging from $M_A\in 6-7~M_{\odot}$ and $M_B\in 5.5-6.3~M_{\odot}$ We find solutions for TIC~470710327$'$ in the range of $11.5-14~M_{\odot}$. We note that the values of the mass of TIC~470710327$'$ are correlated with the mass of TIC~470710327 C due to the imposed light contributions of these objects. The model and observed SED are shown in Fig.~\ref{fig:sed}. We note that this modelling returns a distance in the range of 4-4.5~kpc, whereas the Gaia parallax returns a distance closer to 1~kpc. In order to arrive at a distance close to the Gaia estimate, the reddening would need to be around 2, as opposed to the value in the range of 0.40-0.44 that we find. Although these two distances do not agree, the complexity of this system could contribute to an erroneous astrometric solution, which will likely be improved upon in future Gaia data releases. 

\subsection{Dynamical stability}
\label{subsec:stability}

Even though both the inner and outer orbits of the triple system are relatively compact, the separation between the two orbits ($P_2$/$P_1$ $\approx$ 50) implies that the current configuration is dynamically stable. The stability criterion of \citet{Mar99}, implies long-term dynamical stability for TIC 470710327 for $P_2~\gtrsim$ 18~d.

%Since the system is dynamically stable, it can stay intact for many dynamical timescales, and so one may wonder if we can observe three-body dynamics at play in TIC 470710327. 

Dynamically stable systems can remain intact for many dynamical timescales, giving rise to the possibility of observing three-body dynamics. ZKL cycles \citep{Lid62,Koz62}, for example, would manifest themselves as cyclic changes in the eccentricity of the inner orbit and in the mutual inclination between the inner and outer orbit. However, classical ZKL resonance can only occur in triples with mutual inclinations between 39.2$^{\circ}$ and 140.8$^{\circ}$, meaning that with the mutually inclined of TIC 470710327, of $\approx16.8 _ { -1.4} ^ { +4.2 }$~$^{\circ}$, this effect is unlikely to be significant for the future evolution of the system.

%it is unlikely these will play a role in TIC 470710327 as the classical Lidov-Kozai resonance can only occur in triples with mutual inclinations between 39.2$^{\circ}$ and 140.8$^{\circ}$, whereas the inner and outer orbits of TIC 470710327 are mutually inclined with an angle of $\approx$ 16.8$^{\circ}$ (+4.18,-1.38)$^{\circ}$.

Conversely, the higher-order effects of three-body dynamics could affect the dynamical evolution of this system \citep{Too16}. The eccentric ZKL mechanism \citep[eZKL; see ][for a review]{Nao16} can give rise to more extreme eccentricity cycles for an extended range of inclinations. The magnitude of this effect can be quantified with the use of the octupole parameter \citep{Lit11, Kat11, Tey13,Li14}:

% \begin{equation}
% \epsilon_{\rm oct} \equiv \frac{m_1-m_2}{m_1+m_2} \frac{a_1}{a_2} \frac{e_2}{1-e_2^2}.
% \label{eq:e_oct}
% \end{equation}

\begin{equation}
\epsilon_{\rm oct} \equiv \frac{m_{\mathrm{A}}-m_\mathrm{B}}{m_{\mathrm{A}}+m_\mathrm{B}} \frac{a_1}{a_2} \frac{e_2}{1-e_2^2} \equiv \frac{1-q_1}{1+q_1} \sqrt[\leftroot{-1}\uproot{2}\scriptstyle 3]{\frac{P_1^2}{P_2^2} \left( 1-q_{\rm tot} \right)} \frac{e_2}{1-e_2^2}.
\label{eq:e_oct}
\end{equation}

The eZKL mechanism is expected to be important for the evolution of the system for values of $\epsilon_{\rm oct} \gtrsim 0.001-0.01$, and under the condition that the mass ratio of the inner binary is less than unity. Given the values in Table~\ref{tab:params}, we find $\epsilon_{\rm oct}$=0.001 for $q\approx0.9$ and $\epsilon_{\rm oct}$=0.01 for $q\approx0.3$. Given the morphology of the eclipses in the {\it TESS} data (and considering the substantial diluting third light), initial modelling suggests that the mass ratio of this inner binary is close to unity. Thus, the octupole term is within the range of relevance for the dynamical evolution of this system. The expected timescale of the eZLK cycles are thought to be of order $\tau\approx$ 180~yr / $\sqrt{\epsilon_{\rm oct}}$ \citep{Ant15},  i.e. $\tau\approx$1800-5700~yr. We further investigated these time-scales with simulations using the TrES triple star evolution code \citep{Too16}, and found the time-scales to be consistent with our analytical estimates. However, when including tides and gravitational wave radiation, the maximum amplitude of the eccentricity of the inner binary remains on the order of $e_1\sim0.001$. Detailed constraints on the time-scale and amplitudes of these cycles would require detailed modelling of the inner binary, which remains impossible without full radial-velocity characterisation of the inner components.

%Unfortunately, the octupole parameter depends on the mass ratio of the inner binary which currently remains \textbf{unconstrained}. 

\subsection{Possible formation scenarios}
\label{subsec:formation}

Several theories exist pertaining to the origin of higher order multiple systems. The formation of multiple systems is dependent on fragmentation of the natal material during the formation process. Hierarchical collapse within molecular clouds eventually leads to the formation of dense stellar cores and clumps, whose collapse results in the formation of stars and clusters. Throughout this, equatorial discs are formed through the conservation of angular momentum, which in turn can become involved in the accretion process and the formation of secondary cores. However, there is still debate as to what scale of fragmentation is the main cause of observed massive multiple systems - the fragmentation of the prestellar core or fragmentation of circumstellar discs. In both cases, opacity has a large effect on the initial separations of the systems, which cannot be less than around 10 au due to the opacity limit of fragmentation (\citealt{boss1998}, \citealt{bate1998}). Therefore it is assumed that systems at closer separations than 10 au must have migrated to their observed positions \citep{bate2002}. Some studies present evidence of disc fragmentation creating higher order massive multiple systems during their embedded phases (e.g. \citealt{megeath2005}). Other effects such as dynamical interactions between other companions and discs \citep{eggleton2006} could also cause inner binaries in multiple systems to harden into close orbits. Recent modelling by \citet{oliva2020}, for example, has shown how disc fragments in the discs of massive protostars form through hierarchical fragmentation along spiral arms and migrate to spectroscopic orbits. 

TIC 470710327 presents an interesting puzzle in terms of its formation given our derived geometry of a close binary with a total mass lower than the tertiary star. More massive stars have shorter Kelvin-Helmholtz timescales than those of lower masses, so one would expect that the more massive tertiary was the first star to form. However, if this was the case, it is likely that when this star reached the main sequence (before the inner binary) it would have disrupted the remaining natal material and therefore discontinued the central binary's formation. If the central binary did form first, a more consistent interpretation could be that the inner binary formed through disc fragmentation and the dynamical effects of this binary on the disc could have created a large over-density at large radii. Such over-densities have been shown to occur in circumbinary discs in works such as \citet{price2018}. This over-density could have accreted significant mass, perhaps accelerated by the continuing dynamical effects of the inner binary. In order for this scenario to hold the mass of the disc must have been very large, as the disc fragmentation process would not convert all the disc material into the eventual stars, and the combined mass of the stars in the tertiary system is at least 29 solar masses. However, the largest protostellar discs detected around massive young stellar objects are of order $\sim$10M$_{\odot}$ (e.g. \citealt{johnston2020}, \citealt{frost2021}).

An alternative explanation for the formation of TIC 470710327 is that this system is a result of the fragmentation of the prestellar core as opposed to fragmentation of a disc \citep[e.g., ][]{2007Krumholz}. This is supported by the fact that additional sources surround this triple system in the local region, in the form of  0.5" and $\sim22$" distant companions. With a spurious association probability of $\sim~1\times 10 ^{-5}$ between TIC 470710327 and the 0.5" companion, this closer source is expected to have formed from the same core collapse. While the more distant source at $\sim22$" may not be currently bound to the tertiary system, it could still have come from the same prestellar core. The core collapse scenario circumvents the mass problem described above for disc fragmentation. Should this have occurred, we can assume that the close-binary formed from one collapse event, the tertiary star from another and the distant source from yet another. The close-binary system could have been formed by disc fragmentation as described above, and its hardening into a close-orbit could have been facilitated by dynamical interactions between the 14-17M$_{\odot}$ star and the distant sources from others. Dynamical effects within the collapsing core could also have led to the capture of the 14-17M$_{\odot}$ star by the close-binary system, forming the tertiary we see today. Dedicated radiative hydrodynamical modelling could help disentangle whether core fragmentation, disc fragmentation or a combination of both resulted in the formation of TIC 470710327, whilst repeat observations of all the sources in the region could help distinguish orbits and determine which stars are bound. Furthermore, as there are currently no resolved Gaia parallaxes for TIC~470710327 and TIC~470710327$'$, we cannot concretely rule out the possibility that these two targets are just un-associated nearby objects on the sky.

\subsection{Future evolution}
\label{subsec:evolution}

The observed mass ratio, whereby the tertiary is more massive than the combined mass of the inner binary, implies that the remainder of this system's evolution will be driven by the evolution of the tertiary as this star will be the first to evolve off the main-sequence. Given the expected mass of the tertiary of 14-17 M$_\odot$, star C is expected to fill its Roche lobe at an age of $\sim$~13~Myr, at which point it will start transferring mass towards the inner binary. This type of mass transfer, from an outer star to an inner binary, is expected to occur in $\sim 1\%$ of all triple systems in the Massive Star Catalogue \citep{DeV14,Hamers2021b}. 

%Typically (in 74-87\% of cases) the donor star lies on the asymptotic giant branch (AGB) with a radius of several hundreds of Solar radii, such that the inner binary comfortably fits within the orbit of the outer AGB star \citep{Too20}. However, given the fact that both orbits are relatively small, the tertiary will fill its Roche lobe as it is crossing the Hertzsprung gap i.e. after reaching core hydrogen exhaustion and before the AGB phase of evolution. With a mass ratio of $q=0.776$ between the outer star and the inner binary, the ensuing mass transfer is expected to proceed in a stable manner.

%At an age of $\sim$13Myr, it will fill its Roche lobe and start transferring mass towards the inner binary. This type of mass transfer from an outer star to an inner binary is expected to occur in about $1\%$ of all triple systems in the Massive Star Catalogue \citep{DeV14}. Typically (in 74-87\% of cases) the donor star would lie on the asymptotic giant branch (AGB) with a radius of several hundreds of Solar radii (such that the inner binary comfortably fits within the orbit of the AGB star) \citep{Too20}. However, as TIC 470710327 comprises of two compact orbits, the tertiary will fill its Roche lobe before reaching the AGB and while still crossing the Hertzsprung gap. With a mass ratio of $q\approx 12/15=0.8$ between the outer star and the inner binary, the ensuing mass transfer is expected to proceed in a stable manner (reference?).

The outcome of such a mass transfer phase is, inherently, a hydrodynamical problem \citep{DeV14, Por19}. If the inner binary is compact enough such that the mass transfer stream intersects itself, a circumbinary disk may form. \cite{Lei20} argue that such a scenario leads to preferential accretion to the lowest mass star of the inner binary and therefore favours evolution towards equal mass inner binary stars. The circularisation radius of the mass transfer material \citep{Fra02,Too16} is around $20R_{\odot}$. As the inner binary has an orbital separation of $\sim 10R_{\odot}$, a disk may form, but it is not clear from simple analytical calculation whether that disk would be stable enough to allow for secular accretion. If the mass transfer stream intersects the trajectory of the inner two stars in their orbit around each other, friction may reduce the orbit \citep{DeV14} to lead to a contact system and/or a merger. Such a merger remnant may be considered a blue straggler for two reasons: it would be rejuvenated due to the merger and due to the accretion from the tertiary star.

\begin{figure}
    \centering
    \includegraphics[width=0.45\textwidth]{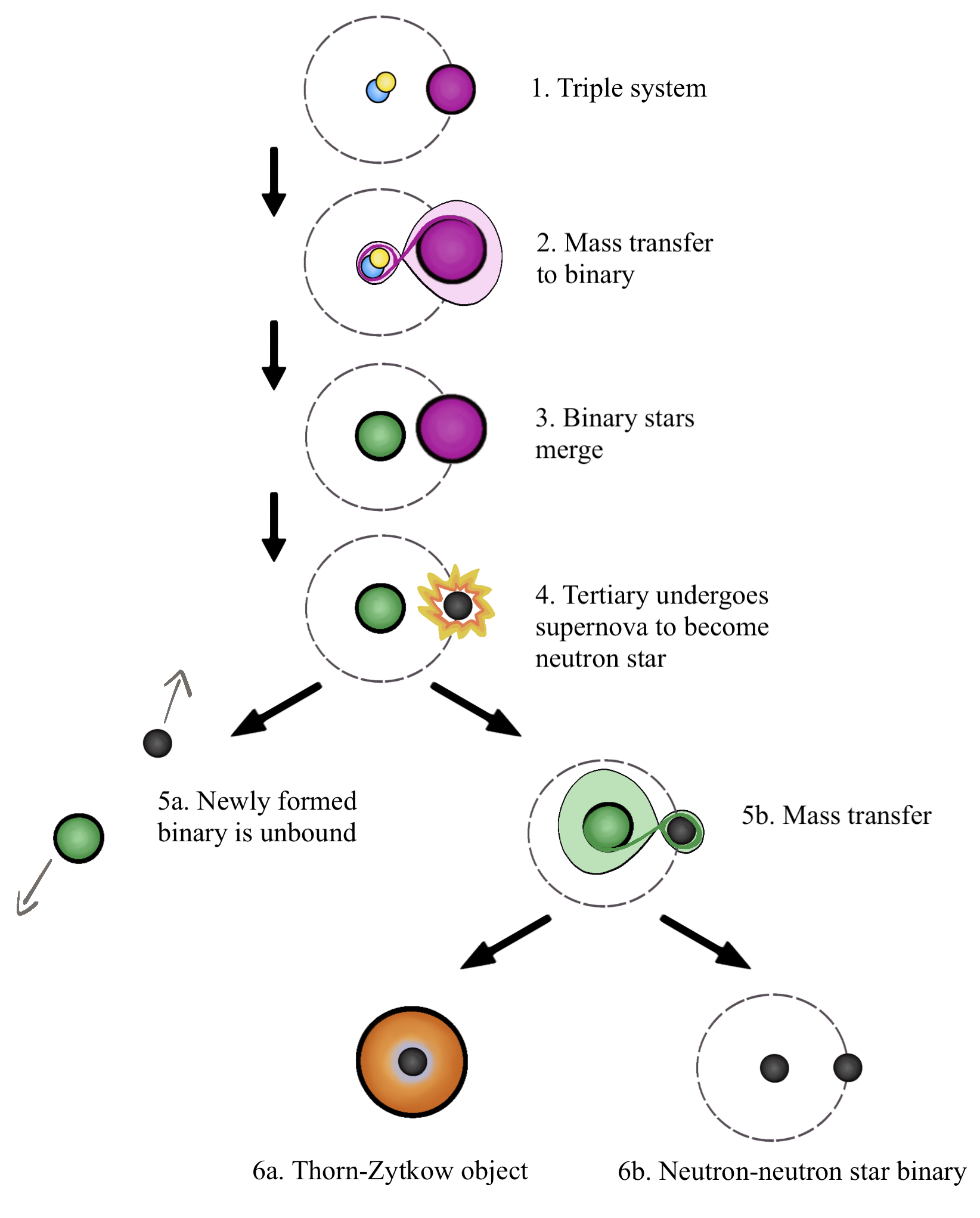}
    \caption{Possible future evolution of TIC 470710327. Relative sizes of the stars and orbits are not to scale. For simplicity, all orbits are depicted as circular.}
    \label{fig:evolution}
\end{figure}

Assuming a merger does take place, the triple would reduce to a binary system (stage 3 in Fig.~\ref{fig:evolution}). After the mass transfer phase ends the merger remnant would have a mass of 12-18 $M_{\odot}$ depending on how efficiently the binary was able to accrete matter. Given the mass of the former donor star it will evolve to become a neutron star. With typical post-mass transfer periods of several hundreds of days (i.e. orbital velocities of around 100 km/s), typical natal kicks from the supernova explosion, with magnitudes of several hundreds of km/s, would unbind the newly formed binary into two single stars \citep{Hob05,Ver17,Igo20}, reducing the multiplicity of the system once more (5a in Fig.~\ref{fig:evolution}). If the orientation of the supernova kick is such that the binary remains intact the binary would likely undergo an additional mass transfer phase when the merger remnant evolves off the main-sequence. Given the large mass ratio of such a binary, the mass transfer would lead to a common-envelope phase \citep{Iva13}. Consequently, this could either lead to a merger and the formation of a Thorne-{\.Z}ytkow object \citep[whereby the neutron star is enclosed by the red giant star, 6a in Fig.~\ref{fig:evolution};][]{Thorne1975,Podsiadlowski1995,Levesque2014,Tabernero2021} or, if the binary survives, experience an ultra-stripped supernova \citep{Tau15} and end up as a double neutron-star (6b in Fig.~\ref{fig:evolution}).

\section{Conclusions}
\label{sec:conclusion}
TIC~470710327 is a compact, hierarchical triple system consisting of a 1.10 d binary containing two B-type stars and an OB-type tertiary on a wide 52 d orbit. The system was initially identified in \textit{TESS} data by citizen scientists taking part in the Planet Hunters TESS project. Using publicly available {\it TESS} data and newly obtained {\sc Hermes} data, we report on the dynamical modelling of the system to reveal a rare configuration wherein the tertiary object in the wide orbit is more massive than the combined mass of the inner binary ($m_{\mathrm{AB}}$=10.9 - 13.2 M$_\odot$, $m_{\mathrm{C}}$=14 - 17 M$_\odot$). This configuration poses several challenges to explain its formation. Given the current mass of the tertiary component, star C is expected to fill its Roche-lobe at an age of $\sim$13 Myr, meaning that the system must necessarily be younger than this age. Considering that the main-sequence lifetime of a 5.5-7 $M_{\odot}$ star is between 40-80 Myr, all of the stars in this system are currently in the main-sequence phase of their evolution. Furthermore, given the values from the dynamical modelling and constrained SED modelling, the primary binary component (star A) will fill between 70-83\% of its Roche-lobe before the tertiary evolves off the main-sequence, while the secondary binary component (star B) will fill between 68-75\% of its Roche-lobe during this time. This means that the binary will not undergo mass transfer before experiencing Roche-lobe overflow from the tertiary.

%As such, this implies that all components are currently in the main-sequence phase of evolution, . Furthermore, given the values from the dynamical modelling and constrained SED modelling, the primary binary component will fill between 70-83\% of its Roche-lobe before the tertiary evolves off the main-sequence, while the secondary binary component will fill between 68-75\% of its Roche-lobe during this time. Exact values depend on the mass ratio of the inner binary.}

%the system is less than $\sim$13 Myr old as the massive tertiary is expected to fill its Roche-lobe at this age. 
Given the compact orbits and the unusually high mass of the tertiary object, we speculate that the future evolution of this system will minimally involve one episode of mass transfer as the massive tertiary evolves across the Hertzsprung gap. Based on its initial mass, the tertiary will likely end its life as a neutron star. Alternatively, depending on the rate and efficiency of the mass transfer to the inner binary, the tertiary could evolve into an intermediate mass stripped star \citep{Gotberg2020}. Should the binary system remain bound after the expected supernova kick (or supernovae kicks), this system could result in a close double neutron star gravitational wave progenitor, or an exotic Thorne-{\.Z}ytkow object. Detection of more systems similar to TIC~470710327 would provide constraints on potential progenitor systems to gravitational wave events. 

In addition to the triple system we report on two nearby stars that significantly contribute to the \textit{TESS} aperture, located at angular separations of 22" (LS I +61 72) and 0.5" (TIC 470710327$'$). Given the field density of stars around TIC 470710327, determined using Gaia eDR3, and the magnitude contrast between the target and the 0.5" companion, we show that the spurious association probability between TIC 470710327 and TIC 470710327$'$ is $\sim~1\times 10 ^{-5}$. These two nearby stars may be the source of the two additional periodic signals ($P_3=9.97~$d and $P_4=4.01$~d) seen in the \textit{TESS} light curve. Additional photometric observations of the nearby stars are needed in order to probe the origins of the $P_3$ and $P_4$ and to determine the true multiplicity of this complex system. 

With further observational characterisation, particularly aimed at characterising the nature of the inner binary, this system stands to become an excellent target to scrutinise simulations of massive star formation and evolution. Future spectroscopic observations that are specifically aimed at detecting the RV variations of both components of the inner binary would allow us to place tighter constraints on the dynamics of the system. In particular, these observations would allow us to probe the eZKL mechanism. Finally, RV characterisation and detailed eclipse modelling of the inner binary would precisely constrain the light contributions of all components of the triple system. With such constraints, derivations of the atmospheric properties, such as T$_{\mathrm{eff}}$, $\log g$, v$_{\mathrm{rot}}$ and L$_{\mathrm{bol}}$ would allow us to further test whether the three stars are coeval.

% The detection of rare systems like TIC 470710327 highlights the potential of space-based photometric mission, such as \textit{TESS}. The cadence, baseline and precision of these data allow for ETV analysis, which, when combined with RV data, allows for the unique determination of...? (tryign to end on somehting positive here :P ) 

\section*{Data availability}
The \textit{TESS} data used within this article are hosted and made publicly available by the Mikulski Archive for Space Telescopes (MAST, \url{http://archive.stsci.edu/tess/}). The {\it TESS} data described here may be obtained from  \url{https://dx.doi.org/10.17909/t9-5z05-k040}. Similarly, the Planet Hunters TESS classifications made by the citizen scientists can be found on the Planet Hunters Analysis Database (PHAD, \url{https://mast.stsci.edu/phad/}), which is also hosted by MAST.

Original {\sc Hermes} spectra, and the newly obtained photometric data are available upon request. 

This work made use of Astropy, a community-developed core Python package for Astronomy \citep{astropy2013}, matplotlib \citep{matplotlib}, pandas \citep{pandas}, NumPy \citep{numpy}, astroquery \citep{ginsburg2019astroquery}, sklearn \citep{pedregosa2011scikit} and {\sc exoplanet} \citep{exoplanet:exoplanet}. 

\section*{Acknowledgements}

We thank the referee for their helpful comments that improved the manuscript. We thank all of the citizen scientists who take part in the Planet Hunters TESS project and who enable the discovery of exciting planet and stellar systems in TESS data. NE also thanks the LSSTC Data Science Fellowship Program, which is funded by LSSTC, NSF Cybertraining Grant number 1829740, the Brinson Foundation, and the Moore Foundation; her participation in the program has benefited this work. Furthermore, NE and SA acknowledge support from the UK Science and Technology Facilities Council (STFC) under grant codes ST/R505006/1 and consolidated grant no. ST/S000488. This work also received funding from the European Research Council (ERC) under the European Union’s Horizon 2020 research and innovation program (Grant agreement No. 865624). CJ has received funding from NOVA, the European Research Council under the European Union's Horizon 2020 research and innovation programme (N$^\circ$670519:MAMSIE), 
and from the Research Foundation Flanders under grant agreement G0A2917N (BlackGEM). ST acknowledge support from the Netherlands Research Council NWO (VENI 639.041.645 grants). PGB was supported by NAWI Graz. SJ acknowledges support from the FWO PhD fellowship under project 11E1721N. KZAC acknowledges support under grant P/308614 of the IAC, which is financed by funds transferred from the Spanish Ministry of Science, Innovation and Universities (MCIU).

Some of the data presented in this paper were obtained from the Mikulski Archive for Space Telescopes (MAST). STScI is operated by the Association of Universities for Research in Astronomy, Inc., under NASA contract NAS5-26555. Support for MAST for non-HST data is provided by the NASA Office of Space Science via grant NNX13AC07G and by other grants and contracts. This paper includes data collected with the TESS mission, obtained from the MAST data archive at the Space Telescope Science Institute (STScI). Funding for the TESS mission is provided by the NASA Explorer Program. STScI is operated by the Association of Universities for Research in Astronomy, Inc., under NASA contract NAS 5–26555.

This work has made use of data from the European Space Agency (ESA) mission {\it Gaia} (\url{https://www.cosmos.esa.int/gaia}), processed by the {\it Gaia} Data Processing and Analysis Consortium (DPAC, \url{https://www.cosmos.esa.int/web/gaia/dpac/consortium}). Funding for the DPAC has been provided by national institutions, in particular the institutionsparticipating in the {\it Gaia} Multilateral Agreement.

Based on observations made with the Mercator Telescope, operated on the island of La Palma by the Flemish Community, at the Spanish Observatorio del Roque de los Muchachos of the Instituto de Astrofísica de Canarias. Based on observations obtained with the HERMES spectrograph on the Mercator telescope, which is supported by the Research Foundation - Flanders (FWO), Belgium, the Research Council of KU Leuven, Belgium, the Fonds National de la Recherche Scientifique (F.R.S.-FNRS), Belgium, the Royal Observatory of Belgium, the Observatoire de Gen{\'e}ve, Switzerland and the Th{\"u}ringer Landessternwarte Tautenburg, Germany.

Finally, NE and CJ wish to thank the Harry Potter franchise for providing us with the in-house nickname for this system of \textit{Fluffy}, inspired by Hagrid’s three-headed dog in Harry Potter and the Philosopher’s Stone.

%%%%%%%%%%%%%%%%%%%%%%%%%%%%%%%%%%%%%%%%%%%%%%%%%%


\bibliographystyle{mnras}
\bibliography{tic470710327} % if your bibtex file is called example.bib

%%%%%%%%%%%%%%%%%%%%%%%%%%%%%%%%%%%%%%%%%%%%%%%%%%

%%%%%%%%%%%%%%%%% APPENDICES %%%%%%%%%%%%%%%%%%%%%

%%%%%%%%%%%%%%%%%%%%%%%%%%%%%%%%%%%%%%%%%%%%%%%%%%
\appendix
\section{Additional Tables}
\label{appendixA}

Summary of the RV observations obtained with the {\sc HERMES} spectrograph (Section~\ref{sec:spec_rv}), ETVs determined from the TESS data (Section~\ref{subsec:ETV} and inputs used for the SED analysis (Section~\ref{subsec:sed}).

\renewcommand{\arraystretch}{1}
\begin{table}
    \centering
    \begin{tabular}{llll}
    Time (BJD - 2457000) & RV (km/s) & RV err  (km/s) & SNR \\
    \hline
    1880.3395            & 0.03      & 7.19           & 45  \\
    1880.3518            & 4.22      & 7.11           & 45  \\
    1883.3908            & 24.25     & 9.00           & 27  \\
    1884.3813            & 46.12     & 5.78           & 59  \\
    1885.3333            & 43.12     & 3.10           & 85  \\
    1886.3368            & 51.28     & 3.63           & 80  \\
    1887.3369            & 42.11     & 2.51           & 91  \\
    1888.3537            & 32.07     & 4.07           & 75  \\
    1889.3285            & 23.70     & 2.72           & 89  \\
    1890.3321            & 18.43     & 3.81           & 78  \\
    1891.3247            & 14.30     & 5.07           & 66  \\
    1892.3270            & 9.70      & 6.93           & 47  \\
    2007.7168            & -36.85    & 2.45           & 92  \\
    2035.6305            & -10.77    & 4.62           & 70  \\
    2038.6952            & 18.08     & 5.63           & 60  \\
    2040.6938            & 29.27     & 2.40           & 92  \\
    2041.6919            & 42.08     & 2.41           & 92  \\
    2042.6432            & 44.82     & 3.16           & 85  \\
    2043.6680            & 47.71     & 3.19           & 84  \\
    2067.7128            & -74.50    & 2.49           & 91  \\
    2069.7160            & -86.61    & 2.48           & 91  \\
    2071.7410            & -95.40    & 5.67           & 60  \\
    2072.7397            & -105.13   & 2.00           & 96  \\
    2127.5437            & -105.09   & 3.46           & 82  \\

    \hline
    \end{tabular}
    \caption{RV observations for TIC 470710327}
    \label{tab:RV}
\end{table}

\begin{table}
    \centering
    \begin{tabular}{llll}
        Cycle  & Predicted linear ephemeris    & O-C & Error \\
         number &  epoch (BJD - 2457000)  &  (mins) & (mins) \\
                \hline
        0      & 1766.2700 & -0.684       & 0.582      \\
        1      & 1767.3747 & 1.147        & 0.583      \\
        2      & 1768.4794 & 1.510         & 0.577      \\
        3      & 1769.5841 & 1.908        & 0.594      \\
        4      & 1770.6887 & 2.403        & 0.592      \\
        5      & 1771.7934 & 1.995        & 0.590       \\
        6      & 1772.8981 & 3.012        & 0.604      \\
        7      & 1774.0028 & 2.351        & 0.603      \\
        11     & 1778.4215 & -1.554       & 0.599      \\
        12     & 1779.5262 & -0.756       & 0.584      \\
        13     & 1780.6309 & 0.378        & 0.581      \\
        14     & 1781.7356 & -0.343       & 0.609      \\
        15     & 1782.8403 & -1.649       & 0.589      \\
        16     & 1783.9450 & -2.364       & 0.602      \\
        17     & 1785.0497 & -1.846       & 0.580       \\
        18     & 1786.1543 & -2.562       & 0.594      \\
        19     & 1787.2590 & -2.571       & 0.598      \\
        24     & 1792.7825 & -4.193       & 0.599      \\
        25     & 1793.8871 & -10.852      & 0.597      \\
        26     & 1794.9918 & -4.375       & 0.644      \\
        27     & 1796.0965 & -4.856       & 0.577      \\
        28     & 1797.2012 & -5.707       & 0.563      \\
        29     & 1798.3059 & -5.709       & 0.575      \\
        30     & 1799.4106 & -5.429       & 0.591      \\
        31     & 1800.5153 & -4.098       & 0.605      \\
        34     & 1803.8293 & -4.444       & 0.581      \\
        35     & 1804.9340 & -4.312       & 0.584      \\
        36     & 1806.0387 & 0.786        & 0.554      \\
        37     & 1807.1434 & -3.490        & 0.568      \\
        38     & 1808.2480 & -3.775       & 0.649      \\
        39     & 1809.3527 & -2.923       & 0.562      \\
        40     & 1810.4574 & -3.381       & 0.563      \\
        41     & 1811.5621 & -1.754       & 0.566      \\
        42     & 1812.6668 & -2.763       & 0.608      \\
        43     & 1813.7715 & -1.453       & 0.587      \\
        172    & 1956.2759 & -1.853       & 0.587      \\
        174    & 1958.4853 & -2.323       & 0.625      \\
        175    & 1959.5900 & -3.002       & 0.590       \\
        176    & 1960.6946 & -1.307       & 0.591      \\
        177    & 1961.7993 & -1.815       & 0.592      \\
        178    & 1962.9040 & -1.221       & 0.594      \\
        179    & 1964.0087 & -3.575       & 0.594      \\
        181    & 1966.2181 & 0.422        & 0.629      \\
        182    & 1967.3228 & -1.289       & 0.599      \\
        184    & 1969.5321 & -0.969       & 0.583      \\
        186    & 1971.7415 & 2.736        & 0.627      \\
        187    & 1972.8462 & 1.779        & 0.615      \\
        188    & 1973.9509 & -3.132       & 0.593      \\
        190    & 1976.1602 & 2.088        & 0.657      \\
        191    & 1977.2649 & 3.076        & 0.611      \\
        192    & 1978.3696 & 4.212        & 0.594      \\
        193    & 1979.4743 & 4.072        & 0.623      \\
        194    & 1980.5790 & 3.980         & 0.597      \\
        195    & 1981.6837 & 3.632        & 0.582      \\
        \hline
    \end{tabular}
    \caption{O-C values for the primary eclipses.}
    \label{tab:ETV_parimary}
\end{table}

\begin{table}
    \centering
    \begin{tabular}{llll}
        Cycle  & Predicted linear ephemeris    & O-C & Error \\
        number &  epoch (BJD - 2457000)  &  (mins) & (mins) \\
         \hline
        0.5    & 1766.8223 & 0.552        & 0.773      \\
        1.5    & 1767.9270 & 0.284        & 0.775      \\
        2.5    & 1769.0317 & 1.908        & 0.789      \\
        3.5    & 1770.1364 & 2.266        & 0.805      \\
        4.5    & 1771.2411 & 4.902        & 0.796      \\
        5.5    & 1772.3458 & 2.988        & 0.786      \\
        6.5    & 1773.4505 & -4.964       & 0.792      \\
        7.5    & 1774.5551 & 3.468        & 0.821      \\
        10.5   & 1777.8692 & 0.403        & 0.850      \\
        11.5   & 1778.9739 & 0.303        & 0.784      \\
        12.5   & 1780.0786 & 0.757        & 0.829      \\
        13.5   & 1781.1833 & -0.089       & 0.791      \\
        14.5   & 1782.2879 & 0.185        & 0.807      \\
        15.5   & 1783.3926 & -1.492       & 0.779      \\
        16.5   & 1784.4973 & -2.271       & 0.794      \\
        18.5   & 1786.7067 & -3.702       & 0.791      \\
        19.5   & 1787.8114 & -2.883       & 0.804      \\
        23.5   & 1792.2301 & -3.806       & 0.763      \\
        24.5   & 1793.3348 & -5.062       & 0.774      \\
        25.5   & 1794.4395 & -4.309       & 0.836      \\
        27.5   & 1796.6489 & -5.181       & 0.763      \\
        28.5   & 1797.7535 & -3.611       & 0.768      \\
        29.5   & 1798.8582 & -4.852       & 0.784      \\
        30.5   & 1799.9629 & -5.658       & 0.776      \\
        31.5   & 1801.0676 & -4.506       & 0.775      \\
        34.5   & 1804.3816 & -5.796       & 0.804      \\
        36.5   & 1806.5910 & -4.506       & 0.809      \\
        37.5   & 1807.6957 & -4.097       & 0.810      \\
        38.5   & 1808.8004 & -3.854       & 0.777      \\
        39.5   & 1809.9051 & -2.949       & 0.835      \\
        40.5   & 1811.0098 & -2.466       & 0.834      \\
        41.5   & 1812.1144 & -2.277       & 0.770      \\
        42.5   & 1813.2191 & -1.383       & 0.811      \\
        173.5  & 1957.9329 & -2.145       & 0.897      \\
        174.5  & 1959.0376 & -1.105       & 0.832      \\
        175.5  & 1960.1423 & -1.108       & 0.793      \\
        176.5  & 1961.2470 & -3.146       & 0.809      \\
        177.5  & 1962.3517 & -4.163       & 0.813      \\
        178.5  & 1963.4564 & -1.103       & 0.814      \\
        179.5  & 1964.5610 & 0.615        & 0.783      \\
        180.5  & 1965.6657 & 1.774        & 0.794      \\
        181.5  & 1966.7704 & -1.163       & 0.793      \\
        182.5  & 1967.8751 & -0.928       & 0.813      \\
        184.5  & 1970.0845 & -0.943       & 0.816      \\
        186.5  & 1972.2938 & 2.103        & 0.830      \\
        187.5  & 1973.3985 & 2.611        & 0.797      \\
        188.5  & 1974.5032 & 1.298        & 0.807      \\
        189.5  & 1975.6079 & 1.833        & 0.832      \\
        190.5  & 1976.7126 & 3.443        & 0.803      \\
        191.5  & 1977.8173 & 7.375        & 0.819      \\
        192.5  & 1978.9220 & 3.657        & 0.796      \\
        193.5  & 1980.0266 & 4.875        & 0.860      \\
        194.5  & 1981.1313 & 4.909        & 0.859      \\
        \hline
    \end{tabular}
    \caption{O-C values for the secondary eclipses.}
    \label{tab:ETV_secondary}
\end{table}

\begin{table*}
%\begin{adjustwidth}{-.5in}{-.5in}
\renewcommand{\arraystretch}{1.1}
    \centering
 % \begin{threeparttable}
    \caption{Photometric observations of TIC~470710327. Data obtained from VizieR (\url{http://vizier.unistra.fr/vizier/sed/}).}
     \begin{tabular}{llcll}
        \hline \hline
        Passband & Effective wavelength [\AA]& Flux [erg s$^{-1}$cm$^{-2}$\AA$^{-1}$]& Magnitude &Ref.\\
        \hline
Johnson:U&3971.00 & 6.52e-13 $\pm$ 3.26e-14 &9.38& a\\
HIP:BT& 4203.01 & 6.40e-13 $\pm$ 1.19e-14 & 9.40&b  \\
HIP:BT& 4203.01 & 6.65e-13 $\pm$ 1.53e-14  & 9.36 &c  \\
Johnson:B &4442.03& 6.17e-13 $\pm$ 1.37e-14 &9.44 & d \\
Johnson:B &4442.03& 6.47e-13 $\pm$ 1.82e-14 &9.39 & e \\
SDSS:g' & 4819.97 &7.88e-13 $\pm$ 7.62e-16 &9.17& f\\
GAIA2:Gbp & 5046.16 & 5.53e-13 $\pm$ 2.35e-15& 9.56 & g\\
HIP:VT &5318.96& 5.30e-13 $\pm$ 9.54e-15 &9.60& b \\
HIP:VT &5318.96& 5.51e-13 $\pm$ 1.27e-14 &9.56& c \\
Johnson:V &5537.05& 5.48e-13 $\pm$ 5.28e-14 &9.57 & d \\
Johnson:V &5537.05& 4.73e-13 $\pm$ 9.78e-15& 9.73 & f \\
GAIA2:G & 6226.21 &3.96e-13 $\pm$ 1.55e-15& 9.92 & g\\
SDSS:r' &  6246.98 &4.70e-13 $\pm$ 1.54e-14 &9.73& f\\
SDSS:i' & 7634.91 & 3.17e-13 $\pm$ 7.19e-16 & 10.16& f\\
GAIA/GAIA2:Grp & 7724.62 & 2.91e-13 $\pm$ 1.51e-15& 10.25 & g\\
2MASS:J &12390.1&9.94e-14 $\pm$ 1.76e-15 &11.42 & h \\
2MASS:H &16494.77 &4.02e-14 $\pm$ 1.10e-15 &12.40&h \\
2MASS:Ks& 21637.85&1.52e-14 $\pm$ 2.56e-16 &13.46&h \\
WISE:W1&33500.11&2.99e-15 $\pm$ 5.34e-17 &15.23 & h \\
WISE:W2 &46000.19&8.56e-16 $\pm$ 1.56e-17 &16.58& i \\
WISE:W3 &115598.23&2.24e-17 $\pm$ 6.73e-19 &20.54& i \\
WISE:W4 &220906.68&2.98e-18 $\pm$ 3.93e-19 &22.73& i \\
        \hline
     \end{tabular}

    \begin{tablenotes}
      \small
      \item \textbf{Notes.} Data from: $^{\text{(a)}}$ \cite{Reed_2003}; $^{\text{(b)}}$ \cite{Fabricius_2002}; $^{\text{(c)}}$ \cite{Bourges_2017}; $^{\text{(d)}}$ \cite{Ammons_2006}; $^{\text{(e)}}$ \cite{Lasker_2006}; $^{\text{(f)}}$ \cite{Henden_2016}; $^{\text{(g)}}$ Gaia DR2 data, see acknowledgements, \citet{Gaia2016,Gaia2018}; $^{\text{(h)}}$ \cite{Zacharias_2012};$^{\text{(i)}}$ \cite{Cutri2013}
    \end{tablenotes}

   \label{table_data_vizier}
%  \end{threeparttable}
%\end{adjustwidth}
\end{table*}

% Don't change these lines
\bsp	% typesetting comment
\label{lastpage}
\end{document}